\documentclass[12pt,letterpaper,english,10pt,letterpaper,english,numberedappendix,apj]{emulateapj}
\usepackage[T1]{fontenc}
\usepackage[latin9]{inputenc}
\usepackage{amssymb}
\usepackage{graphicx}
\usepackage{esint}

\makeatletter

\usepackage{esint}
\usepackage{eucal}

\usepackage{bm}
\usepackage{mathrsfs}

\usepackage{natbib}

\shorttitle{Focal Plane Wavefront Sensing using AO Speckles}
\shortauthors{Codona and Kenworthy}

\makeatother

\usepackage{babel}
\begin{document}

\title{Focal Plane Wavefront Sensing using Residual Adaptive Optics Speckles}

\author{Johanan L. Codona}

\affil{Steward Observatory, University of Arizona, Tucson, AZ 85721}

\author{Matthew Kenworthy}

\affil{Leiden Observatory, Leiden University, Leiden, The Netherlands}

\affil{Steward Observatory, University of Arizona, Tucson, AZ 85721}

\email{jlcodona@gmail.com}
\begin{abstract}
Optical imperfections, misalignments, aberrations, and even dust can
significantly limit sensitivity in high-contrast imaging systems such
as coronagraphs. An upstream deformable mirror (DM) in the pupil can
be used to correct or compensate for these flaws, either to enhance
Strehl ratio or suppress residual coronagraphic halo. Measurement
of the phase and amplitude of the starlight halo at the science camera
is essential for determining the DM shape that compensates for any
non-common-path (NCP) wavefront errors. Using DM displacement ripples
to create a series of probe and anti-halo speckles in the focal plane
has been proposed for space-based coronagraphs and successfully demonstrated
in the lab. We present the theory and first on-sky demonstration of
a technique to measure the complex halo using the rapidly-changing
residual atmospheric speckles at the 6.5~m MMT telescope using the
Clio mid-IR camera. The AO system's wavefront sensor (WFS) measurements
are used to estimate the residual wavefront, allowing us to approximately
compute the rapidly-evolving phase and amplitude of speckle halo.
When combined with relatively-short, synchronized science camera images,
the complex speckle estimates can be used to interferometrically analyze
the images, leading to an estimate of the static diffraction halo
with NCP effects included. In an operational system, this information
could be collected continuously and used to iteratively correct quasi-static
NCP errors or suppress imperfect coronagraphic halos. 
\end{abstract}


\section{Introduction.\label{sec:intro}}

Extrasolar planets (ESPs) are expected to be $10^{6}$--$10^{9}$
times fainter than their host stars in the thermal infrared. Solar
system analogues will have planetary orbital distributions within
one arcsecond of their host star \citep{2011PhDT.........3N}, where
they are masked by diffracted and scattered starlight. Without a coronagraph,
most ESPs will appear to be several decades fainter than the telescope
diffraction pattern, characterized by its \emph{point spread function}
(PSF). A coronagraph suppresses the diffraction structures in the
PSF halo, but wavefront aberrations, alignment errors, and dust or
other transmission flaws within the telescope scatter starlight into
the suppressed region, overwhelming the faint science signal from
the ESP. Ground-based telescopes rely on adaptive optics (AO) to correct
the atmospheric wavefront distortions, but even a high degree of correction
leaves random, rapidly-changing, residual speckles that are several
decades brighter than the ESPs. Ideally, long exposures should average
out the speckle noise as $1/\sqrt{t}$ to a level where the planet
becomes detectable against a smooth background. Unfortunately, any
unsuppressed halo is coherently modulated by the AO speckles, increasing
the speckle noise \citep{pinned,Aime04}, and AO processing lag can
cause the residual speckles to be both brighter and persist longer
than they otherwise might \citep{1990JOSAA...7.1224F,not_nature2002}.
The most serious features in the scattered starlight are \emph{quasi-static
speckles} (QSS) caused by \emph{non-common-path} (NCP) aberrations
\citep{Racine99,Hinkley07}, which are not known \emph{a priori},
change on timescales from minutes to hours, and average slowly---if
at all \citep{2005SPIE.5903..170M,Sivaramakrishnan06}. These halo
features determine the practical detection limit for ESPs. 

A number of techniques have been developed to address QSS and the
effect of slowly-changing PSFs over time. These techniques include
\emph{Angular Differential Imaging} (ADI) \citep{Marois06,Schneider03},
\emph{Simultaneous Differential Imaging} (SDI) \citep{Biller07,Racine99,Marois00},
and \emph{Local Optimization of Combined Images} (LOCI) \citep{Lafreniere07},
all with various degrees of success. These techniques all post-process
the science camera data after observation time, and therefore can
only achieve incoherent gains. AO systems include a \emph{deformable
mirror} (DM) that can dynamically modify the incoming wavefront, allowing
wavefront correction to be introduced pre-detection, with the potential
for coherent improvements in the signal gain. In an AO system, after
encountering the DM, part of the starlight is diverted into a \emph{wavefront
sensor} (WFS), where the wavefront shape is measured and processed
by a computer, sending updated compensating shape commands to the
DM. The resulting flattened starlight wavefront may optionally be
used to feed a coronagraph or imaged directly onto a science camera.
In either case, NCP aberrations that occur downstream from the AO
system are only evident in the final science images. This can manifest
as distorted and fainter images or coronagraphically-suppressed halos
that are spoiled by scattered starlight. But if instead of correcting
the wavefront from the vantage of the WFS, we can try to improve the
science image \citep{1995PASP..107..386M} by applying bias offsets
to the DM, we can obtain coherent gains that are significantly better
than those achieved post-detection. Suppressing the halo also suppresses
the interference or ``pinned'' speckle noise \citep{pinned}, with
a corresponding improvement in sensitivity. The problem is determining
the correct biases to apply.

Uniquely improving the PSF (or the Strehl ratio) or suppressing undesired
halo in a coronagraph using information from the science camera, requires
more than just intensity measurements: it requires phase as well.
A full complex halo measurement can be inverse Fourier-transformed
to estimate the NCP optical flaws, interpreted under the paradigm
of Fourier optics. The Strehl ratio can be improved by biasing the
DM with the opposite of the computed pupil wavefront. Coronagraphic
halo suppression can be improved by measuring the residual halo's
complex amplitude and creating a set of ``anti-speckles'' by adding
ripples to the DM to suppress them \citep{CodonaAngel_ApJ_2004}.
It is possible to build a modified Lyot coronagraph that uses the
blocked starlight to form a reference beam \citep{not_nature2002,CodonaAngel_ApJ_2004}
in a focal plane interferometer. However, measuring the halo with
the science camera requires that the reference beam be included, spoiling
the image unless exposures are multiplexed between interferometer
and science. The two modes will have have similar exposure times if
we are to measure faint static speckles that are of the same level
of brightness as a planet, and the corresponding loss of science exposure
time leads to a loss in sensitivity.

Another approach is to use the upstream DM to create weak probe speckles
in the focal plane with controllable phase. The speckles coherently
add to the underlying halo and interfere, modulating the intensity
which is then recorded by the science camera \citep{2006ApJ...638..488B,Giveon07}.
By using three or more probe speckle phases it is possible to measure
the complex amplitude of the static halo under the probe speckle.
This method works very well and is the basis for the extraordinary
laboratory halo suppression results at the JPL \emph{High-Contrast
Imaging Testbed} (HCIT) \citep{Trauger07}, designed to simulate a
very stable space-based environment. Implementing this on the ground
has to also contend with the presence of residual AO speckles, as
well as a much less stable environment where at least the quasi-static
speckles are changing on the timescale of a few minutes \citep{Hinkley07,martinez2012speckle},
forcing the the probe speckles to be brighter in order to speed the
measurements. Regardless of how the complex halo measurements are
made, the same anti-halo servo algorithms can be applied or, if the
the entire complex halo is available, Strehl ratio improvements can
be achieved at the science camera. The downside of using a speckle
probing method is the same as the coronagraphic focal plane interferometer:
it puts probe light into the science image, forcing science and wavefront
sensing to be multiplexed, reducing sensitivity.

In this paper we present and demonstrate a new technique for measuring
the complex halo in the focal plane using only the residual AO speckles.
Since the speckles are already present in the field of view, no extra
light is introduced during the measurement, hence there is no need
to reduce science exposure time. Since the AO speckles vary rapidly,
measurements of both quasi-static speckles and other halo structure
can be obtained fast enough to implement focal plane servos with update
rates of several times per minute. Our method does not attempt to
control the residual AO speckles, but only uses the WFS telemetry
to monitor and characterize them as they change. The method requires
no new hardware and should be implementable on virtually any AO system.
The only requirements are that the science camera be capable of sufficiently
short exposures ($\lesssim$ 30 ms) to adequately ``freeze'' the
speckles, and that the camera and WFS data be accurately synchronized.
In an earlier paper \citep{Codona08}, we proposed an algorithm called
``Phase Sorting Interferometry'' (PSI), which binned and sorted
the pixels based on the computed speckle phase. That result gave the
static halo's phase, which is the information needed to properly place
a DM ripple in an anti-halo servo. In the present paper we greatly
simplify the analysis into a set of statistical formulae that can
easily be computed in real time. If the halo is estimated over only
a limited search area, as it would be with a coronagraph, the results
can be used to suppress the residual halo. If the halo is estimated
more globally, including the PSF core, the result can also be used
to estimate and correct the NCP aberrations, improving Strehl ratio
and PSF quality at the science camera. 

Our technique is most closely related to the random phase diversity
techniques employed by \citet{lee1997diagnosing} and more recently
by \citet{frazin2013}. In those papers the focal plane intensity
is written as a nonlinear functional of the unknown NCP aberrations,
as well as the known residual AO wavefront error. The NCP error is
then estimated from a synchronized set of focal plane images, simultaneous
WFS measurements, and a nonlinear algorithm. In contrast, our approach
is primarily to create the functional equivalent of a focal plane
interferometer that measures the mean halo field. Once known, the
linear relationship between the focal plane and pupil plane fields
can be used to estimate the mean pupil field. If desired, the aberrated
wavefront may then be estimated by computing the complex argument
of the mean pupil field. Our approach never requires a nonlinear algorithm
to estimate the pupil wavefront, the phase nonlinearity being completely
contained within the calculation of the pupil field argument.

In Section 2 we present the theory of interferometry using a known-but-random
probe field. In section 3 we simulate an AO system with NCP aberrations
and apply the theory to the simulated data set. We demonstrate the
robustness of the estimated complex halo and derived NCP aberrations
by adding a wide range of noise in the science images. In Section
4 we use Shack-Hartmann WFS data from the MMTO 6.5~m telescope and
short-exposure images from the Clio mid-infrared camera to estimate
the complex halo. By Fourier transform, we estimate the complex pupil
field including the effect of subsequent NCP aberrations. Using a
7th magnitude test star, reasonably consistent NCP aberration estimates
were available in a few seconds. More stable estimates would required
longer integrations, but an update to a possible NCP-correcting servo
would be possible every 30 seconds or more.

\section{Theory\label{sec:theory}}

\newcommand{\doublehat}[1]{
\settoheight{\dhatheight}{\ensuremath{\hat{#1}}}
\addtolength{\dhatheight}{-0.35ex}
\hat{\vphantom{\rule{1pt}{\dhatheight}}
\smash{\hat{#1}}}}

\global\long\def\Ltwo{\mathbb{L}_{2}}

\global\long\def\xx{\mathit{\mathbf{x}}}

\global\long\def\xxv{\mathit{\mathbf{x}}}

\global\long\def\kk{\bm{\kappa}}

\global\long\def\qq{\mathbf{q}}

\global\long\def\vv{\mathbf{v}}

\global\long\def\ss{\mathbf{s}}

\global\long\def\kkv{\bm{\kappa}}

\global\long\def\alphav{\bm{\alpha}}
\global\long\def\betav{\bm{\beta}}

\global\long\def\thv{\bm{\theta}}

\global\long\def\xiv{\bm{\xi}}

\global\long\def\eps{\epsilon}

\global\long\def\wind{\mathbf{v}_{wind}}

\global\long\def\DIAM{\mathrm{D}}

\global\long\def\lact{l_{actuators}}

\global\long\def\louter{L_{0}}

\global\long\def\lrecon{l_{recon}}

\global\long\def\ld{\lambda/D}

\global\long\def\phase{\phi}

\global\long\def\phat{\phi_{DM}}

\global\long\def\SF{D_{\phase}}
\global\long\def\SFhat{D_{fit}}
\global\long\def\SFhhat{D_{DM}}

\global\long\def\phasebar{\bar{\phi}}
 \global\long\def\delphase{\delta\phase}
 \global\long\def\varphase{\eta}

\global\long\def\dzed{\delta z}
\global\long\def\dd{\mathrm{d}}

\global\long\def\gain{\mathit{g}}

\global\long\def\field{\psi}
\global\long\def\dfield{\delta\field}

\global\long\def\halo{\Psi}
\global\long\def\halobar{\bar{\Psi}}
\global\long\def\halos{\delta\halo}

\global\long\def\FF{\mathcal{F}}

\global\long\def\SR{\mathscr{S}}
\global\long\def\Strehl{\mathscr{S}}

\global\long\def\pupil{\Pi}
\global\long\def\dpupil{\delta\Pi}

\global\long\def\PSF{\Phi}

\global\long\def\PSFsp{\Phi_{speckles}}
\global\long\def\OTF{\mathcal{O}}

\global\long\def\e{e}

\global\long\def\REAL{\mathbb{R}}
\global\long\def\COMPLEX{\mathbb{C}}
\global\long\def\INTS{\mathbb{Z}}

\global\long\def\SUBJECT{\bm{\Psi}}
\global\long\def\REF{\bm{\psi}}

\global\long\def\SS{\Psi}
\global\long\def\RR{\psi}
\global\long\def\SSmag{\left|\Psi\right|}
\global\long\def\RRmag{\left|\psi\right|}

\global\long\def\QQ{\mathcal{Q}}

\global\long\def\MCF{\Gamma}

\global\long\def\tnaught{\tau_{0}}
\global\long\def\tlag{\tau_{lag}}
\global\long\def\tAO{\tau_{0'}}
\global\long\def\ts{\tau_{\mbox{speckle}}}
\global\long\def\tphase{\tau_{\varphase}}
\global\long\def\tsci{\mathsf{t}_{sci}}

\global\long\def\twfs{\mathsf{t}_{wfs}}

\global\long\def\Tcube{{\cal T}}

\global\long\def\PSpeckles{\mathfrak{S}}

\global\long\def\MGF{{\cal M}}

\global\long\def\lambdasci{\lambda_{science}}

\global\long\def\Texp{\mathsf{t}_{science}}
\global\long\def\SLOPES{\bm{\Theta}}

\global\long\def\RECONao{\boldsymbol{W}_{AO}}
\global\long\def\RECON{\boldsymbol{W}}

\global\long\def\RECONspeckles{\boldsymbol{W}_{residual}}

\global\long\def\DISPLACEMENTS{\bm{z}}

\global\long\def\delz{\delta\bm{z}}
\global\long\def\PHIsp{\Phi_{speckles}}
\global\long\def\VARsp{\sigma_{speckles}^{2}}
\global\long\def\SIGMAsp{\sigma_{speckles}}

\global\long\def\lao{\ell_{AO}}
\global\long\def\phaseref{\phase_{ref}}
\global\long\def\pfield{\Upsilon}
\global\long\def\pphase{\varphi}

We begin by considering a set of short-exposure images of a star,
acquired while the telescope's AO system is delivering diffraction-limited
images with a clearly visible PSF core. The images consist of a static
halo (the diffraction pattern of the telescope) and a cloud of rapidly-changing
speckles. The speckles are significantly fainter than they would be
be without the AO system, but they are still present and contribute
scattered starlight out to the radius of the seeing disk and beyond.
The science camera is unable to directly measure the incident electrical
field's phase, only the intensity of the static halo as it is modulated
and coherently interfered with by changing speckle cloud. To see the
speckles and their interference clearly, the individual exposures
must be short enough that the speckles do not change much in phase
or amplitude. The residual speckles arise from several sources: wavefront
correction errors from noise in the wavefront sensing, limitations
or errors in the wavefront reconstruction, AO servo errors (e.g.~loop
gain), fitting errors due to the deformable mirror, and processing
lag. Of these, lag error speckles are the most dynamic since they
are related to the wind along the telescope's line of sight. The speckle
cloud typically appears as diffraction-limited noise, with individual
speckles having widths of $\sim\ld$. But the halo also contains speckles
and possibly deformations and extended regions of scattered starlight
with much longer timescales. These include the quasi-static speckles
which confusion-limit sensitivity in high-contrast detection. For
our discussion, we consider these more slowly-changing features to
be a part of the ``static'' halo, and by measuring its complex amplitude
we can provide an adaptive anti-halo servo algorithm with the information
required to correct aberrations or suppress unwanted halo and maximize
sensitivity. 

Whatever the cause of the rapidly-changing uncorrected wavefront error,
and certainly in the case of lag error, the WFS residuals are non-zero
and can still be measured. While the continuing goal for high-end
AO systems is to reduce the residual wavefront error to below the
WFS noise floor, we are not there yet. It is true that some of the
residual error cannot be seen by the WFS, but in the current generation
of AO systems there is always statistically-significant measurable
wavefront error while the AO system is in closed loop. We will assume
that we are in that situation. Any unsensed but varying residual will
appear to us as unexplained noise in the image data. Changes that
are averaged over during the science exposures will appear as a loss
of coherence. Classic errors such as waffle are not a problem with
the MMT since the hexapolar actuator pattern is incommensurate with
the WFS subapertures. It might be an issue when implementing this
method on other AO systems however.

Even though the Taylor ``frozen flow'' hypothesis (i.e. the approximation
that turbulent motion of the atmospheric irregularities may be ignored
as they are carried past the telescope's line-of-sight by the wind)
may be nearly true for the incident uncorrected wavefront aberrations,
the residual AO wavefront error does not behave the same way. Consider
a typical low-altitude wind that carries atmospheric irregularities
across the 6.5~m MMT aperture in about 0.5~s. High altitude winds
are typically much faster and cross the aperture in about 100 -- 200~ms.
In either case the irregularities are not likely to significantly
rearrange during the pupil crossing. Turbulence within the observatory
dome or in the vicinity of the telescope optics are different and
significant rearrangement is likely. The AO system estimates the wavefront
error and approximately removes it after a processing delay. It does
this using a wavefront error-suppressing servo with an adjustable
gain factor. As a particular wavefront pattern crosses the pupil,
the AO system iteratively measures the residuals and adds corrections
to the DM. Even for rapidly-moving wavefront patterns, the AO system
has many tens to hundreds of iterations to suppress it. As a result,
the residual pattern is usually incrementally altered as it moves
across the pupil, becoming progressively more uncorrelated with its
earlier configuration as it moves. The effect of an incorrect servo
gain is that any pattern will not be completely removed in the next
iteration. This repeats even while the initial pattern is being carried
across the pupil, leaving a residual wavefront pattern that exponentially
decays in-place with a time constant that depends on how far the gain
is from the ``correct'' value. Thus the residual wavefront error
is typically dominated by two or possibly three parts: a set of changing
wavefront patterns that appear and disappear with a time scale dependent
on the servo gain, and one or two wind-driven patterns that decorrelate
much faster than turbulent rearrangement would suggest. Meanwhile
in the focal plane, the wind-driven lag error forms a wide plume of
speckles that are brighter near the star. In addition, as the wavefront
errors move across the pupil, the speckles' complex phasors rotate
at a speed which increases linearly with distance from the star projected
along the direction of the wind \citep{not_nature2002}. Eventually,
the phase wrapping is so fast that the speckle phase changes significantly
during a single science camera exposure, causing the speckle and the
static halo to appear to lose coherence in that interference between
them has less contrast. Superimposed on this systematic effect is
a random evolution of the speckles that depends on how long it takes
to replace the pupil-plane aberrations with a statistically unrelated
set. This is usually a fraction of $D/\left\Vert \wind\right\Vert $.
These effects, along with the general fading of the speckle halo with
angular distance from the star creates a practical outer radius for
using speckles as a halo measuring tool. Fortunately, the generally
brighter and slower speckles nearer the star allow us to measure the
aberrations that affect telescope or coronagraphic performance exactly
where improvement is needed most.

\begin{figure}
\begin{centering}
\includegraphics[width=1\columnwidth]{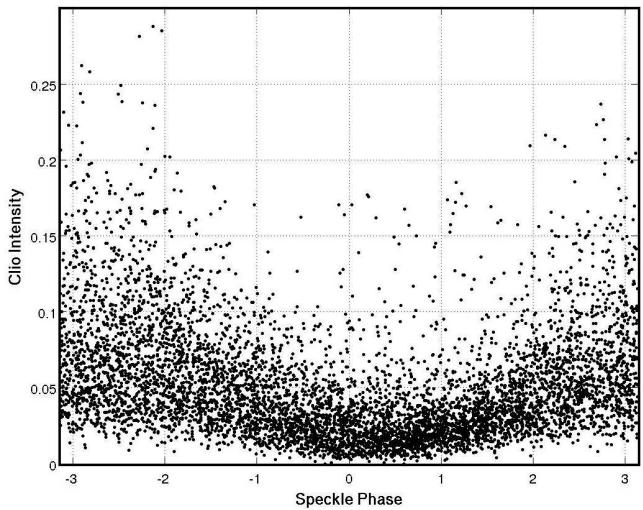}
\par\end{centering}

\caption{\label{fig:MMT-example-data}Complex halo measurements using MMT data
for a single pixel in the first Airy ring. The science camera intensity
is plotted against the speckle phase computed from the WFS measurements.
The speckles vary randomly in phase and amplitude while the underlying
halo is presumed to be steady. When the speckle phase is the same
as the halo, the intensity recorded by the science camera is increased.
When the phase is $180^{\circ}$ off from the halo, the intensity
is reduced. If the speckle amplitude were constant, the intensity
would vary sinusoidally with speckle phase. The randomly-varying amplitude
causes the sinusoid to become noisy, but the mean behavior is the
same. Just judging by eye we can see that the underlying halo has
a phase of about -2.5 rad in this pixel. Statistical analysis allows
for much more precise estimation of the halo phase. Intensity scatter
vs.~intensity swing allows us to estimate the underlying halo amplitude.
See Sec.~\ref{sec:observations} for details.}

\end{figure}

In isolation, the phase of a speckle is unobservable with the science
camera unless the speckle coherently interferes with the static halo.
But the cloud of rapidly-changing residual AO speckles do constantly
interfere with the halo, making an intricate pattern of intensity
fluctuations that can be captured by the science camera. Using the
WFS telemetry and an idealized model for the telescope that ignores
NCP errors, we can compute the phase and amplitude of the speckles
at the science camera. Comparing these calculated speckles with the
intensity recorded by the camera allows us to use the speckles as
interferometric probes, enabling estimation of the phase and amplitude
of the static halo. This relies on a simple concept: when the speckles
are \emph{in phase} with the underlying halo, the intensity is greater,
and when they are anti-phased the combined intensity is diminished.
If we were to plot the observed halo brightness against the speckle
phase (Fig.~\ref{fig:MMT-example-data}), the maximum in the light
curve corresponds to the phase of the halo at that point. In addition
to the phase, an estimate of the static halo's amplitude can be found
by comparing various  statistics computed from the science images
and the computed speckles.

We begin by deriving the interferometry equations needed to use an
ensemble of known random (in the sense of \emph{measured but uncontrolled})
reference beams to estimate a static subject beam. We then apply the
resulting equations to the problem of random speckles and a static
halo. Since we do not have any\emph{ a priori} knowledge about the
NCP aberrations, our calculations of the complex speckle field can
only be approximate. However, the phase of the speckle peaks and their
locations are able to be computed rather robustly and the effects
of the unknown NCP errors will only enter as ``speckles on the speckles''
which will appear as a small amount of noise in our measurement.

\subsection{\label{sub:Phase-Shifting-Interferometry}Traditional Interferometry
Theory}

Since we have the ability to create speckles in the science camera's
focal plane by impressing a ripple on the DM, we could simply cycle
through three or more ripple phases (i.e. by changing the ripple placement
relative to the pupil) and determine which placement reinforces or
suppresses the static halo under the resulting speckle \citep[see ][]{Giveon07,Kenworthy06}.
The analysis is straightforward and equivalent to the common interferometry
technique of using a phase-shifted reference beam to probe an unknown
subject beam. In that generalized case the subject beam under test
is mixed with a phase-shifted reference beam and a camera is used
to capture the resulting interference pattern or ``interferogram.''
Using residual AO speckles is similar to using intentionally created
speckles, except that they vary randomly in both phase and amplitude
and are beyond our control. Even though the evolution is random, we
can still measure the residual wavefront and determine the speckle
amplitude and phase by calculation. The temporal evolution of the
speckles is only important in that they might change during a single
science camera exposure, smearing the parameters that we would like
to use in interpreting the resulting image. Otherwise, the random
speckles can be thought of as instantaneous samples drawn from an
ensemble described by statistical moments. Interferometry with a random
collection of reference beams (or speckles) looks different than the
conventional analysis with carefully selected phases and a constant
amplitude, but the principles are the same. Therefore, we will start
by reviewing phase-shift interferometry, but use terminology that
extends cleanly to our random speckle case. Both the traditional
and generalized theory described below can be applied to measuring
fields in any optical plane or physical context, but we will only
consider its application in the focal plane in this paper. The traditional
goal is to measure the phase of a constant \emph{subject} \emph{beam}
$\SS$ by mixing it with a constant \emph{reference beam} $\RR$ that
is stepped through a set of known phase offsets. At each phase offset
the camera is used to capture an \emph{intensity interferogram} that
is the magnitude-squared of the sum of the subject and phase-shifted
reference beam fields. In our treatment we consider a single pixel
of the camera and the fields are complex constants with phases $\phase_{s}=\arg\SS$
and $\phase_{r}=\arg\RR$ (where for $\psi=\left|\psi\right|\exp\left\{ i\phi\right\} $,
$\phi=\arg\psi\equiv\Im\ln\psi$). The reference beam is phase-shifted
by an extra amount $\varphase\in\{\varphase_{n}\}$. To solve for
the subject beam, we need at least three phase shifts in order to
uniquely determine the phase. Even though everything here is constant
or preset, we will use angle bracket notation $\left\langle \cdot\right\rangle $
to indicate averaging over an appropriate or indicated set. This will
simplify the notational transition to randomly-varying reference beams
easier. We require that the phase shifts are selected such that 
\begin{equation}
\left\langle \exp\left\{ i\varphase_{n}\right\} \right\rangle \equiv\frac{1}{N}\sum_{n=1}^{N}\e^{i\varphase_{n}}=0.\label{eq:zeroMeanUnitPhasors}
\end{equation}
If this condition is not met, the non-zero-mean reference beam will
leave a residual that cannot be distinguished from being part of the
subject beam. Our analysis will therefore produce the correct value
biased by the mean of the reference beam. We also require that 
\begin{equation}
\left\langle \exp\left\{ 2i\varphase_{n}\right\} \right\rangle =0\label{eq:doublePhaseScrambledUnit}
\end{equation}
so that we may easily process the interferograms. The interferograms
record the intensity of the interfering fields at each of the applied
phase shifts, giving 
\begin{eqnarray}
I_{n} & = & \left|\SS+\RR\e^{i\varphase_{n}}\right|^{2}\label{eq:MixerCohSum}\\
 & = & \SSmag^{2}+\RRmag^{2}+\SS\RR^{*}\e^{-i\varphase_{n}}+\SS^{*}\RR\e^{i\varphase_{n}}\label{eq:MixerComplexNotation}\\
 & = & \SSmag^{2}+\RRmag^{2}+2\RRmag\SSmag\cos(\phase_{s}-\phase_{r}-\varphase_{n}).\label{eq:cohFringes}
\end{eqnarray}
The expanded real form in Eq.~\ref{eq:cohFringes} is the most familiar,
but we prefer Eq.~\ref{eq:MixerComplexNotation} because the complex
algebra is simpler. We proceed by assigning each interferogram to
the shifted complex phase of the reference beam and averaging over
the phase shifts
\begin{equation}
\left\langle I_{n}\e^{i\varphase_{n}}\right\rangle =\left(\SSmag^{2}+\RRmag^{2}\right)\left\langle \e^{i\varphase_{n}}\right\rangle +\SS\RR^{*}+\SS^{*}\RR\left\langle \e^{2i\varphase_{n}}\right\rangle .\label{eq:normalInterferometerProc}
\end{equation}
Making use of our constraints Eqs.~\ref{eq:zeroMeanUnitPhasors}
and \ref{eq:doublePhaseScrambledUnit}, we find that the mean intensity
term averages to zero, and one of the two varying interference terms
is ``stabilized'' while the other averages to zero by construction.
The single surviving term is the product of the subject and conjugated
reference beam 
\begin{equation}
\left\langle I_{n}\e^{i\varphase_{n}}\right\rangle =\SS\RR^{*}.\label{eq:stabilizedInterferograms}
\end{equation}
Taking the imaginary part of the log of Eq.~\ref{eq:stabilizedInterferograms},
we find the subject beam phase to be 
\begin{equation}
\phase_{s}=\arg\left\langle I_{n}\e^{i\varphase_{n}}\right\rangle +\phase_{r}.\label{eq:subjectPhase}
\end{equation}

In many interferometric applications, the only desired quantity is
the subject beam phase, in which case we are done. In cases where
the subject beam amplitude is also desired, it can be found by simply
blocking the reference beam and measuring the non-interfering intensity.
But since we cannot ``turn off'' the speckles, along with the fact
that the speckle cloud's shape and amplitude fluctuates with the wind
and other external influences and may not be accurately known, we
will now find the subject beam amplitude assuming the reference beam
amplitude is unknown. From Eq.~\ref{eq:MixerComplexNotation} we
find the average intensity as 
\begin{equation}
\left\langle I\right\rangle =\SSmag^{2}+\RRmag^{2}.\label{eq:IncoherentIntensity}
\end{equation}
The magnitude-squared of Eq. \ref{eq:stabilizedInterferograms} gives
us the product of beam intensities 
\begin{equation}
\left|\left\langle I_{n}\e^{i\varphase_{n}}\right\rangle \right|^{2}=\SSmag^{2}\RRmag^{2}.\label{eq:stabMag}
\end{equation}
Note that $\SS$ and $\RR$ are interchangeable in both of these equations.
Using Eqs.~\ref{eq:stabMag} and substituting in Eq.~\ref{eq:IncoherentIntensity}
and solving for the subject beam intensity, 
\begin{equation}
\SSmag^{2}=\frac{\left\langle I_{n}\right\rangle }{2}\pm\frac{1}{2}\sqrt{\left\langle I_{n}\right\rangle ^{2}-4\left|\left\langle I_{n}\e^{i\varphase_{n}}\right\rangle \right|^{2}},\label{eq:subjectIntensity}
\end{equation}
we find two solutions. This is due to the subject-reference ambiguity
built into the statement of the problem. Which beam was physically
phase shifted in our measurements is lost when looking at the intensity,
and we are left with two solutions: one being the subject beam intensity
and the other the reference beam intensity.

\subsection{\label{sub:Random-Ref-Interferometry}Interferometry with a known
random reference beam}

The analysis in Sec.~\ref{sub:Phase-Shifting-Interferometry} is
very rigid in its reference beam variation, the main purpose being
to simplify the mathematics. But we can also use a randomly-changing
reference beam to analyze a static subject field $\SS$ in each pixel
of a set of images in a manner very similar to Sec.~\ref{sub:Phase-Shifting-Interferometry}.
Being ``random'' in this context means not being controlled or predictable.
``Random'' does not imply that the reference beam is unknown, since
we will always assume that at least its phase is known. If we know
both the reference beam's phase \emph{and} amplitude corresponding
to a series of images, the problem is very similar to the phase-shifted
reference beam case above. Unlike that analysis however, which was
unable to tell the difference between subject and reference beam phase
and amplitudes, the varying reference beam allows us to unambiguously
solve for both the subject beam since it is constant while the reference
beam changes. If all we know about the reference beam is its phase,
we can still estimate the subject beam's phase by using the known
reference beam phase to stabilize one of the fluctuating terms in
Eq.~\ref{eq:MixerComplexNotation}. Even though the amplitude is
changing, this allows a reliable statistical estimate of what reference
phase maximizes the intensity (Fig.~\ref{fig:MMT-example-data}).
When the intensity is maximized, the subject and reference phases
are the same. If we also know the reference beam amplitudes, or even
if we just know the rms speckle amplitude, we can determine the subject
beam amplitude. Since our reference beam is determined by calculation
from the WFS measurements and does not depend on, for example, the
flux of the star (Sec.~\ref{sub:CSpeckles-from-WFS}), we also have
a scale factor between reference and subject beams that needs to be
determined from the data's statistics. 

In Sec.~\ref{sub:Phase-Shifting-Interferometry}, we typically had
3 or 4 images with known phase offsets. For the random reference problem,
we imagine having a synchronized pair of relatively large data sets:
one of complex reference beam fields and the other of the corresponding
intensities. We treat each synchronized collection of images and fields
as a set with statistics determined from the available data. We make
no assumptions about underlying probability distributions.

The total field is the sum of the subject and reference fields, with
the reference beam changing without the need for a separate phase
offset 
\begin{equation}
\halo_{total}=\halo+\field.
\end{equation}
The subject field $\halo$ is assumed to be constant over the data
set, while the reference beam $\field$ changes in some uncontrolled
yet measured or otherwise known way. The first assumption is that
the reference beam field has a zero mean, functionally corresponding
to the reference phase shift requirement in Eq.~\ref{eq:zeroMeanUnitPhasors},
but allowing for varying amplitudes, 
\begin{equation}
\left\langle \psi\right\rangle =0.\label{eq:zeroMean}
\end{equation}
If we were analyzing an ensemble of data, this would have some more
absolute meaning. But since it is just the mean over the available
data, perhaps spanning just a few seconds, the mean will have some
random residual. This residual will appear to be part of the static
subject beam, affecting the derived subject beam as estimation error.
If the reference beam mean is non-zero, we can subtract it from the
complex reference values and proceed using Eq.~\ref{eq:zeroMean}
\begin{equation}
\halo_{total}=\left(\halo+\left\langle \field\right\rangle \right)+\left(\field-\left\langle \field\right\rangle \right).
\end{equation}
The analysis will now yield the biased result $\left(\halo+\left\langle \field\right\rangle \right)$,
from which the means may be subtracted at the end of the calculation.
For simplicity, we will continue to write the reference field with
the suppressed mean as $\field$ which makes Eq.~\ref{eq:zeroMean}
true by construction. In addition to Eq.~\ref{eq:zeroMean}, we note
that the mean square (not conjugated) of the field becomes negligible
in large data sets,
\begin{equation}
\left\langle \psi^{2}\right\rangle \rightarrow0.\label{eq:zeroSqMean}
\end{equation}
This is the analog of Eq.~\ref{eq:doublePhaseScrambledUnit} and
cannot be independently forced like the mean. In our case where the
speckles (our reference beam) are each the sum of many independent
patches of approximately-flattened field across the pupil, the complex
speckle field tends toward a Gaussian distribution by the central
limit theorem (App.~\ref{app:additive_speckles}). Note that for
a zero-mean Gaussian random field, the scatter in estimates of $\left|\left\langle \psi^{2}\right\rangle \right|\propto1/\sqrt{N}$
where $N$ is the number of statistically-independent speckles in
each data set (App.~\ref{app:GRV}). The same is true of any moment
$\left\langle \psi^{n}\left(\psi^{\star}\right)^{m}\right\rangle $
where $n\ne m$. We only require this general result for one more
term, where the scatter in estimates of $\left\langle \psi\psi^{\star}\psi\right\rangle \rightarrow0$
as $1/\sqrt{N}$ which is likely true so long as the speckle intensity
and phase are uncorrelated (see App.~\ref{app:GRV}).

We start as before with a simple intensity model, leaving out any
questions of partial coherence or intensity-dependent noise. We simply
treat the static halo as a complex constant coherently interfering
with the changing but known reference field. The resulting intensity
is given by 
\begin{eqnarray}
I & = & \left|\SS+\RR\right|^{2}\label{eq:CohSum}\\
I & = & \SSmag^{2}+\left|\psi\right|^{2}+\SS\RR^{*}+\SS^{*}\RR.\label{eq:CohSumExpanded}
\end{eqnarray}
Taking the average of Eq.~\ref{eq:CohSumExpanded} and invoking $\left\langle \psi\right\rangle =0$,
we find the subject beam intensity is 
\begin{equation}
\SSmag^{2}=\langle I\rangle-\langle\left|\psi\right|^{2}\rangle\equiv\Phi_{I}-\PHIsp.\label{eq:meanIntensity}
\end{equation}
To facilitate our use of these results in later sections, we have
also written the result in terms of physical measurables: $\Phi_{I}\equiv\langle I\rangle$
is the mean PSF recorded by the science camera and $\PHIsp\equiv\langle\psi\psi^{\star}\rangle$
is the average intensity of the halo of speckles if they were recorded
on their own. Note that while the intensity expressed in Eq.~\ref{eq:CohSumExpanded}
is real, each of the conjugated cross terms $\SS\RR^{*}$ and $\SS^{*}\RR$
rotate in opposite directions and individually have zero means. We
can stabilize the first of the fluctuating terms in Eq.~\ref{eq:CohSumExpanded}
by multiplying through by $\psi$, maintaining a stable phase while
causing the other term to move twice as fast in phase. Upon averaging,
the stabilized term survives and the others drop out giving
\begin{equation}
\left\langle \psi I\right\rangle =\langle\left|\psi\right|^{2}\rangle\SS\equiv\PHIsp\SS.\label{eq:stabilized}
\end{equation}
In writing this equation, we also made use of our assumption that
$\langle\left|\psi\right|^{2}\psi\rangle\rightarrow0$ even though
it will not be precisely zero for a shorter data set, but will be
in a larger data set. Since $\langle\left|\psi\right|^{2}\rangle\in\REAL$,
we can immediately determine the phase of the subject beam as 
\begin{equation}
\arg\Psi=\arg\left\langle \psi I\right\rangle ,\label{eq:subjectPhaseStabilized}
\end{equation}
which is the robust result we described in \citet{Codona08}. Since
the reference beam intensity must be non-zero in order to make any
measurement at all, we can simply write the subject field as 
\begin{equation}
\Psi=\frac{\left\langle \psi I\right\rangle }{\langle\left|\psi\right|^{2}\rangle}\equiv\frac{\left\langle \psi I\right\rangle }{\PHIsp}.\label{eq:RSI-halo-fullyCoh-noGain}
\end{equation}

\subsection{\label{sub:ABetterModel}Including an unknown scale factor}

The subject beam field $\Psi$ is the static diffraction halo in the
science camera focal plane. The reference beam field $\field$ is
the speckle field caused by the changing residual wavefront errors
after the AO system has ``corrected'' the atmospherically-distorted
incident field. The intensity $I$ measured by the science camera
is proportional to the magnitude-squared of the sum of the static
halo and speckle fields, but with the additional factors of incident
flux and exposure time, etc. The speckle field is computed from WFS
slopes with possible (modal) gain errors in the wavefront reconstruction,
and generic assumptions about amplitude. Converting these fields to
actual images require all of the aforementioned factors such as starlight
flux. All of these influences can be included in an unknown scale
factor $\gain$ that may also be a function of focal plane position.
The relationships between $\Psi$, $\field$, and $I$ depend on getting
this scale factor right, so we will add it to the model and estimate
it from the data. We introduce the scale factor into Eq.~\ref{eq:CohSum}
with the modification
\begin{equation}
I=\left|\SS+g\RR\right|^{2}\label{eq:betterIModel}
\end{equation}
where $g\in\REAL$ since we assume our phase estimates are accurate.
Following our derivation as before, we find the mean intensity
\begin{equation}
\Phi_{I}=\left|\SS\right|^{2}+g^{2}\PHIsp\label{eq:betterMean}
\end{equation}
and the stabilized intensity 
\begin{equation}
\left\langle \RR I\right\rangle =g\SS\Phi_{speckles}.\label{eq:betterStabilized}
\end{equation}
To determine the scale factor before we know $\Psi$, we adjust it
until the intensity variance based on the calculated speckles matches
that of the actual intensity images. Using Eq.~\ref{eq:betterIModel}
to compute $\sigma_{I}^{2}\equiv\langle\left(I-\left\langle I\right\rangle \right)^{2}\rangle$,
we find
\begin{equation}
\sigma_{I}^{2}=2g^{2}\left|\SS\right|^{2}\PHIsp+g^{4}\sigma_{speckles}^{2}\label{eq:betterVariance}
\end{equation}
where $\sigma_{speckles}^{2}=\langle(\left|\RR\right|^{2}-\langle\left|\RR\right|^{2}\rangle)^{2}\rangle$
is the speckle-only intensity variance without halo interference.
Using the magnitude-squared of Eq.~\ref{eq:betterStabilized} to
replace $g^{2}\left|\SS\right|^{2}$ with measured quantities in Eq.~\ref{eq:betterVariance},
we find
\begin{equation}
g=\left[\frac{\sigma_{I}^{2}}{\VARsp}-\frac{2\left|\left\langle \RR I\right\rangle \right|^{2}}{\PHIsp\VARsp}\right]^{1/4}.\label{eq:scale-factor}
\end{equation}
Using this result in Eq.~\ref{eq:betterStabilized} allows us to
write the halo as 
\begin{equation}
\Psi=\frac{\left\langle \RR I\right\rangle }{g\PHIsp}.\label{eq:TheAnswer}
\end{equation}
This is the diffraction halo at the science camera, including any
non-common-path aberrations. While it is obvious that the intensity
imaged by the science camera must first have any background noise
removed, it is important to note that the background noise variance
should also be estimated and removed from $\sigma_{I}^{2}$ before
using Eq.~\ref{eq:TheAnswer}. We will also find that the region
near the center of the halo (very low spatial frequencies) tends to
be biased low due to physical reasons not included in our isolated
pixel analysis. Since the result is still useful without increasing
the complexity of the model to eliminate the bias, we will use these
results as they stand.

\subsection{Visibility}

The traditional interference fringe ``visibility'' measures the
intensity modulation caused by changing the reference beam phase normalized
by the mean intensity \citep[Sec. 7.5]{born1999poe}. It is usually
written in terms of the max and min intensities as 
\begin{eqnarray}
V & = & \frac{I_{max}-I_{min}}{I_{max}+I_{min}}.\label{eq:Vis-classic}
\end{eqnarray}
Since the standard deviation of the cosine is $\sqrt{2}$, we can
use Eq.~\ref{eq:cohFringes} to rewrite the conventional definition
of visibility as
\begin{equation}
V=\frac{\sqrt{2}\sigma_{I}}{\left\langle I\right\rangle }.\label{eq:Vis-classic-statistical}
\end{equation}
In our random reference case, the appropriate portion of the intensity
variance is the term in Eq.~\ref{eq:betterVariance} that is caused
by interference ($2g^{2}\left|\SS\right|^{2}\PHIsp$) rather than
intrinsic variations in the reference beam intensity ($g^{4}\sigma_{speckles}^{2}$).
This allows us to write 
\[
V=\frac{2g\left|\SS\right|\sqrt{\PHIsp}}{\left\langle I\right\rangle }.
\]
Using Eq.~\ref{eq:TheAnswer} to replace $\left|\SS\right|$, and
the notational association of $\left\langle I\right\rangle =\Phi_{I}$,
the formula for the visibility becomes 
\begin{equation}
V=\frac{2\left|\left\langle \RR I\right\rangle \right|}{\Phi_{I}\,\sqrt{\PHIsp}}.\label{eq:visibility}
\end{equation}

\section{Simulation}

\label{sec:Simulation}In this section we demonstrate our technique
with a realistic simulation based on the MMTO 6.5~m telescope and
AO system, with some arbitrary NCP errors added. Using the simulation
results, we apply our random reference interferometric formulae to
estimate the static complex halo and informally explore the effect
of measurement noise. Finally, we Fourier transform the complex halo
back into the pupil plane to estimate the NCP aberrations. We find
that this method underestimates the very lowest spatial frequencies
of the pupil field, requiring a simple correction to recover the proper
static wavefront aberration. The results are adequate for building
either an anti-halo servo for a coronagraph or improving the Strehl
ratio at the science camera by compensating for NCP aberrations. Finally,
we discuss the degradation of results caused by speckle halo evolution
during the science camera exposures and how to estimate its effect
from real data.

\begin{figure}
\begin{centering}
\includegraphics[width=1\columnwidth]{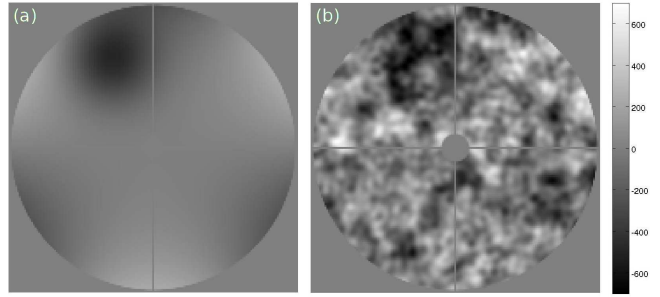}\caption{\label{fig:Sim-NCP}(a) The simulated non-common-path error masked
by the pupil. The modeled aberration includes trefoil and a Gaussian
indentation. The peak-to-valley aberration is 725 nm. (b) The NCP
error added to a realization of the residual AO residual WFE. }

\par\end{centering}

\end{figure}

We used a simple lagged spatial filtering model to simulate the post-AO
wavefront at the MMT. We first synthesized a $2048\times2048$ grid
with 4~cm spacing of Kolmogorov wavefront displacements scaled to
give a Fried length $r_{0}$ of 15~cm in V band and an outer scale
of 30~m. The effect of AO correction was simulated by applying a
Gaussian smoothing to the wavefront and subtracting the result from
the original wavefront. This approximates the effect of applying a
mode-limited reconstructor to the DM. The width of the smoothing kernel
was adjusted to make the rms residual wavefront fitting error (WFE)
250 nm. The effect of wind and processing lag was introduced by laterally
shifting the smoothed correction by $\delta\xx=v_{wind}\tau_{lag}$.
For convenience, the lateral shift was selected to be one grid pixel
of 4~cm. A total of 5 phase screens were synthesized and a set of
residual wavefronts was selected from each by using a sub-grid of
size $D=6.5$~m at $0.85D$ spacings giving a $14\times14$ sample
of slightly-overlapping wavefronts. Each scintillation-free pupil
field instance was computed using $\exp\left\{ 2\pi i\delz/\lambda\right\} $
and multiplied by a 6.5~m pupil mask $\pupil(\xx)$ with a 10\% central
obstruction. The wavelength $\lambda$ was set to 3.8 $\mu$m (L'
band). In addition to the residual AO wavefront error, the camera
images were presumed to include an additional non-common-path (NCP)
wavefront error described by Zernike trefoil ($\delz=Z_{3}^{3}(2r/D,\theta)\,\lambda/16$)
and an additional Gaussian indentation centered on $\xx_{0}=(-1,2)$
(meters) with depth $\lambda/8$ (max mirror figure error of 237.5
nm) with a $1/\e$ radius of 1~m (Fig.~\ref{fig:Sim-NCP}).
 The PSFs were computed using 2-D FFTs padded to $2048\times2048$
samples and taking the magnitude squared to find the intensity. The
resulting images were converted to array counts assuming a mean total
of 157,000 counts per exposure (typical of a magnitude 7.5 star at
L' band, bandwidth of 0.65 $\mu$m, MMT aperture, 50\% QE, and a throughput
of 67\%). Gaussian noise was added to the images to simulate the variance
caused by different amounts of thermal and sky background noise. The
complex speckle halo was computed without the NCP aberrations, but
with the same pupil mask. This takes the place of the WFS-based halo
calculation in the actual observations. The intentional neglect of
the NCP WFE in the halo calculation emulates the fact that we cannot
see the downstream NCP errors when working from the WFS measurements.
This ``bootstrapping'' technique of the algorithm should have minimal
effect on the speckle phases so long as the Strehl ratio is high enough
to have a clear peak above the speckles. (We tested the effect of
this approximation by including the NCP errors in the speckle calculation,
with an effect on the estimated wavefront error of 8~nm rms or about
0.013 radians rms.) The resulting complex halo has speckles with the
ideal PSF's sidelobes, as opposed to the actual speckle sidelobes
which would show the trefoil, etc. The computed speckles are otherwise
in the correct locations and have the correct phase relationship to
the PSF core. After subtracting off the mean complex halo, the remaining
speckles have mostly the correct complex values near their maximum
amplitudes, where the effect of the simplified sidelobes can be ignored.

\begin{figure}
\centering{}\includegraphics[width=6cm]{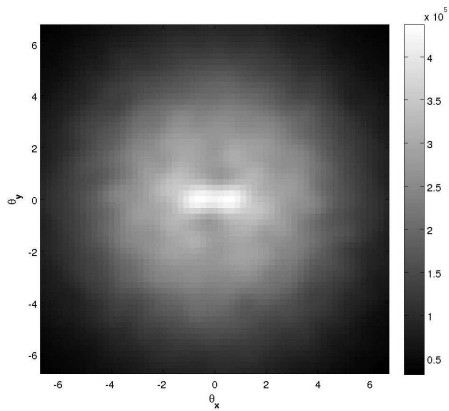}\caption{\label{fig:Sim-SpeckleHaloMeanPower}The mean speckle halo intensity
averaged over the simulated data set. The residual AO fitting error
is isotropic, while the lag error driven by the wind ($\wind\tlag=4$~cm)
is responsible for the horizontal plume of speckles near the center.}

\end{figure}

In a calculation of this sort, it is possible to know and control
the piston component of the wavefront, resulting in absolute phase
control or to have a global phase reference. This is not the case
in reality however. The wavefront sensor only measures wavefront slopes
and is therefore blind to constant phase offsets. The mean intensity
distribution of the speckles, $\PHIsp$, has power and phase fluctuations
all the way in to the center of the PSF. We multiply the complex halo
by a unit-amplitude phasor to adjust the halo's peak value to zero
phase and reference the speckles to the current halo peak, allowing
them to be compared. Ideally, we would want to reference the halo
phase to the core of the static halo, but it is only practical to
use the phase of the full computed complex halo. This will slightly
bias the reference phase and introduce a fluctuating phase error to
all points in the halo. This phase error is zero-mean and becomes
less significant at higher Strehl ratios like those in our simulation.
Our phase referencing procedure does not change either the speckles'
intensity or variance, which is why we correctly see speckle power
all the way to the center of the PSF (Fig.~\ref{fig:Sim-SpeckleHaloMeanPower}).
When we process real data, we also remove tip-tilt from the WFS measurements
and the science images, which has the effect of further removing speckle
power to the next order about the center of the PSFs. Again, this
is not detrimental to the measurement of $\Psi$ away from the center,
but it will bias the lowest spatial frequencies and become an issue
when we attempt to estimate the pupil field from our focal plane results. 

Once the complex halo was computed for all pupil wavefronts and all
of the peaks were referenced to zero phase, the complex mean was computed
and subtracted, leaving only the complex speckles $\psi$.
 The mean speckle intensity is shown in Fig.~\ref{fig:Sim-SpeckleHaloMeanPower}.
The figure clearly shows the broad fitting error cloud and a horizontal
plume of speckles caused by lag error and the wind. There is no evidence
of the main diffraction pattern in the speckle halo, since the appearance
of aberrations in the simulation is independent of position relative
to the pupil. To explore the robustness of the result, zero-mean Gaussian
noise was added to the simulated CCD images to simulate observations
of fainter target stars. The introduced noise sigmas are 0, 10, 100,
and 1000 counts per pixel. 

\begin{figure}
\centering{}\includegraphics[width=6cm]{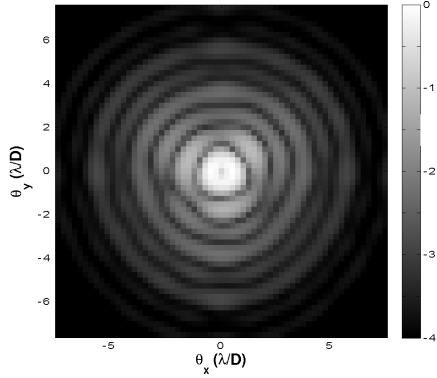}\caption{\label{fig:Sim-PSF-Sigma2}The variance of the simulated science camera
images $\sigma_{I}^{2}$ displayed on a 4-decade logarithmic scale.
This figure only includes intensity variations caused by the changing
speckles, with no background or photon noise. The science image variance
includes the product of the speckle halo and the intensity of the
static halo (Eq.~\ref{eq:betterVariance}). There is a reduction
in the variance at the center as a consequence of the anti-Hermitian
symmetry of speckles and is not included in our analysis model.}

\end{figure}

In order to use the random reference interferometry results, we need
to compute various statistics from the simulated fields and images.
The mean speckle field is zero by construction, leaving any actual
non-zero means folded into the static halo as estimation error.
\begin{figure}
\centering{}\includegraphics[width=6cm]{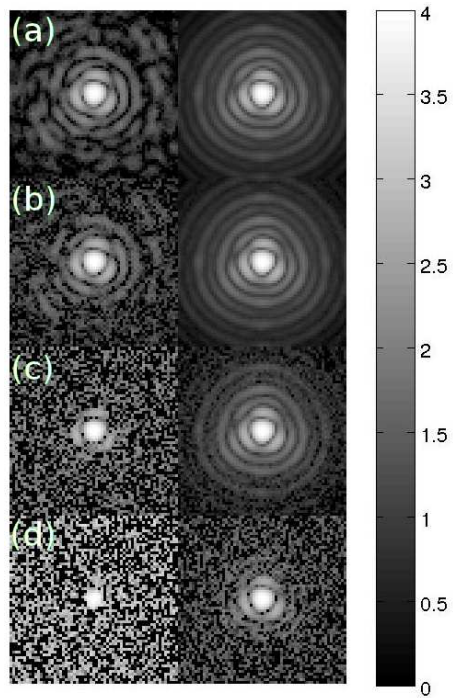}\caption{\label{fig:Simulation-Example-PSFs}Simulated science camera images:
single frame (left) and data set averages (right). The images are
constructed from the computed PSFs with photon noise and added background
noise. The starlight is set to an average of 157,000 counts per exposure,
with added background noise sigmas of (a) 0, (b) 10, (c) 100, and
(d) 1000 counts per pixel. The PSF images are $15\ld$ on each side
with $0.25\ld$ pixels. The simulation dataset included 980 images.
The PSFs on the left are individual 20 ms images while those on the
right are averaged over the 980 images in the entire data set (a total
exposure time of 19.6~s).}

\end{figure}
 The simulation generates two data cubes: a set of simulated science
camera images with all aberrations included and actual counts per
pixel, and a data cube of complex halos that, in reality, would have
been computed from the WFS measurements and a simplified model of
the telescope without the NCP aberrations. The mean of the complex
halo data cube is an estimate of the static halo, but without the
NCP distortions we wish to measure. This mean is subtracted from the
complex halo cube leaving only the complex speckle field. While the
computed speckle phases and relative amplitudes values are meaningful,
the units are arbitrary and cannot be compared directly with the photon
counts in the science camera image cube. So we include a scale factor
$g$ (first introduced in Eq.~\ref{eq:betterIModel}) as a constant
that depends on the flux and takes care of the units. The average
of the science images gives $\Phi_{I}$ (Fig.~\ref{fig:Simulation-Example-PSFs}),
the average of the magnitude-squared speckle fields (the ``speckle
intensities'') gives $\PHIsp$ (Fig.~\ref{fig:Sim-SpeckleHaloMeanPower}),
the variance of the science images is $\sigma_{I}^{2}$ (Fig.~\ref{fig:Sim-PSF-Sigma2})
and the variance of the speckle intensities is $\VARsp$.
 Since the two datasets are computed from the same grids, they are
automatically commensurate and properly aligned. We multiply the science
image and complex speckle field data cubes element-by-element and
average over the 960 frames to compute $\left\langle \RR I\right\rangle $
from which we can derive the static complex halo $\Psi$ using Eq.~\ref{eq:TheAnswer}
(Fig.~\ref{fig:Sim-PSI-Vis}). 
\begin{figure}
\centering{}\includegraphics[width=1\columnwidth]{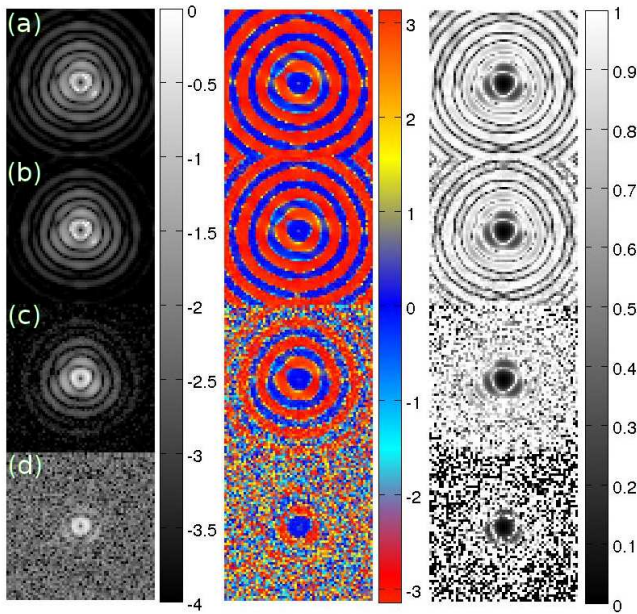}\caption{\label{fig:Sim-PSI-Vis}The computed static halo and visibility at
noise levels listed in Fig.~\ref{fig:Simulation-Example-PSFs}. The
leftmost column is the base-10 log of the static halo amplitude. The
levels are normalized to a point in the first Airy ring. The center
column is the computed phase of the static halo, and the rightmost
column is the visibility (lighter is better). Note the missing intensity
at the very center of the halo in the left column images. This is
due to the anti-Hermitian symmetry of the speckles, an effect that
is not included in our interferometry analysis.}

\end{figure}
 Note that the static halo has a lower amplitude than it should at
the very center of the function. This is caused by a subtle difference
between our interferometric analysis and the symmetries of complex
speckles. We assumed that an arbitrary random complex speckle could
be added to the static halo anywhere during the interferometric analysis.
This is true everywhere in the focal plane except at the center of
the star image. The reason is that a small amplitude wavefront ripple
in the pupil plane not only diffracts starlight from the PSF core
into a speckle at the expected position, but creates another speckle
on the opposite side of the star. Referenced to the starlight in the
PSF core, the two speckles have different phases such that the real
part of the field is antisymmetric, while the imaginary part is symmetric.
This is called \emph{anti-Hermitian }symmetry and is a well-known
feature of faint speckles. Each speckle is able to take on an arbitrary
complex value, subject to energy constraints, and the speckle on the
other side of the star will follow according to its anti-Hermitian
symmetry. This non-local correlation is not an issue in our analysis
because the two correlated speckles are usually so far apart from
each other that they have no mutual effect. However, a very low spatial
frequency aberration, or even small residual tip-tilts, cause speckle
pairs which appear close enough together that they can significantly
interfere. When they do, the antisymmetric real part tends to cancel
and the sum becomes only due to the symmetric imaginary part. Since
we have constructed the phase reference to be the PSF core, ideally
of the static halo center, the center of the static halo is real.
Adding a small speckle near the core approaches purely imaginary and
has the effect of a phase shift. This does not change the intensity
of the PSF core, only the phase, therefore the intensity variance
near the core should drop relative to where the speckle pairs don't
overlap (Fig.~\ref{fig:Sim-PSF-Sigma2}).

\begin{figure}
\centering{}\includegraphics[width=6cm]{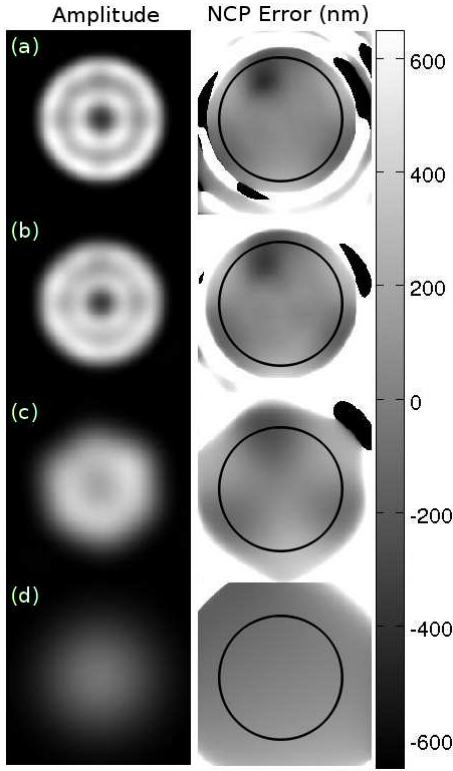}\caption{\label{fig:Sim-PField}Fourier transforming the static halo $\Psi$
computed from Eq.~\ref{eq:PSI-LowFreq} gives an estimate of the
static pupil field $\pfield$ for the same noise levels listed in
Fig.~\ref{fig:Simulation-Example-PSFs}. The left column is the amplitude
of the estimated pupil field and the right column shows the pupil
phase scaled to wavefront displacement in nm. As the noise level increases,
high angles are lost first, leaving a progressively smoothed pupil
wavefront estimate.}

\end{figure}

If we use our results in an anti-halo servo designed to suppress unwanted
residual halo in a coronagraph, the low spatial frequency power deficit
in Eq.~\ref{eq:TheAnswer} does not affect us. However, if we wish
to estimate the wavefront in the pupil plane, as we might for increasing
the Strehl ratio in the science images by correcting the NCP aberrations,
or for computing a more accurate numerical PSF for post-detection
processing of the science images, we would like to recover the low
spatial frequencies as well as the higher ones. We can still do this
by reconsidering how we compute $\left|\Psi\right|^{2}$. Instead
of using Eq.~\ref{eq:betterStabilized} to compute $\Psi$, we compute
the halo phase $\arg{\Psi}$, which is what we found in Eq.~\ref{eq:subjectPhaseStabilized}.
We then estimate the static halo intensity from Eq.~\ref{eq:betterMean},
which does not suffer from the anti-Hermitian speckle effect. To satisfy
the requirement that $\left|\Psi\right|\in\REAL$ and $\left|\Psi\right|^{2}>0$,
we use a $\max$ function, giving a better estimate of $\Psi$ as
\begin{equation}
\Psi=\sqrt{\max\left\{ \Phi_{I}-g^{2}\PHIsp,0\right\} }\e^{i\arg\left\langle \psi I\right\rangle }.\label{eq:PSI-LowFreq}
\end{equation}
We computed $\Psi$ using this equation and Fourier transformed back
into the pupil plane to compare with the initial NCP aberrations.
The results for a set of noise levels is shown in Fig.~\ref{fig:Sim-PField}.
 For each noise level, the pupil field amplitude is shown on the left
and the phase converted back to nanometers of wavefront displacement
is shown on the right, which should be compared with the introduced
NCP aberrations. The spatial resolution is limited because the SNR
drops off with increasing radius. As the noise level is increased,
the estimated pupil field and wavefront becomes progressively smoother.
The simulation is somewhat unrealistic in that the speckles were not
derived from a Nyquist-limited WFS measurements and the wavefront
itself has properties that are statistically ``cleaner'' than real
data with residual wavefront errors that are not tied to the pupil
or actuator locations. Nevertheless, the results validate the technique
and show us what to expect with real data.

\subsection{\label{sub:Exposure-PCoh}Effects of optical bandwidth and exposure
time}

In the preceding discussions, we assumed that the evolving halo field
was monochromatic and instantaneously sampled in time, with no blurring
or phase averaging due to speckle evolution during the exposure. However,
these effects can be important in practice. The incident wavefront
at a single point in the pupil experiences an rms phase change of
1 radian after a time $\tnaught$ defined by the phase structure function
$\SF(\wind\tnaught)=1$ , where $\wind$ is the characteristic wind
velocity projected onto the pupil plane. This time scale determines
how fast WFS exposures must be made as well as actuator and AO servo
bandwidths. However, the speckles do not depend only on the field
at a point, but the field across the entire pupil. Moving the pattern
by a small distance, or even possibly by $r_{0}$, leaves most of
the uncorrected incident wavefront unchanged, simply translated. This
causes speckles to remain in the same place, but with a phase shift
depending on the location of the given speckle in the focal plane
relative to the wind. AO residual speckles are more complicated because
they are the result of the initial aberrations and the spatial and
temporal limitations of the AO system. 

For atmospheres with visible $r_{0}$ values of $\sim15$~cm and
winds of $\sim20$~m/s, $\tnaught\gtrsim2$~ms. An AO wavefront
sensor typically has readout speeds of 500 -- 2000 frames/s, allowing
the wavefront to be estimated at that frame rate. Limited photon flux
degrades the wavefront estimate accuracy with increasing frame rate,
as does decreasing star brightness or increasing background noise
\citep{1994OSAJ...11..925S}. The MMT AO system is currently at the
low end of the frame rate range, but still fast enough to capture
the low spatial frequency wavefront evolution without significant
averaging. Science cameras are usually used with much slower frame
rates, although capabilities of 30 to 50 fps are not uncommon. Broader
optical filter bandwidths pass more light, enabling us to make better
use of higher frame rates. However, the broader filter bands also
increase chromatic effects that can lead to confusion in the analysis.

Our MMT L-Band images have optical bandwidths of about 20\% (0.7~$\mu$m
bandwidth centered on 3.7~$\mu$m). This introduces chromatic radial
smearing where both diffraction halo and speckle features appear at
points proportionally farther from the star with increasing wavelength.
A practical operational constraint is to not allow speckles and other
halo features to radially blur into each other, beyond which they
may become confusion-limited. The limits implied by this constraint
can be estimated by noting that the typical monochromatic speckle
size and separation are both $\sim\lambda_{0}/D$ ($\lambda_{0}$
is a characteristic wavelength in the detected band), regardless of
their location relative to the star. At a radius $\theta$, the radial
smearing is $\left(\delta\lambda/\lambda_{0}\right)\theta$. The non-overlap
constraint gives us the condition 
\begin{equation}
\theta<\left(\frac{\lambda_{0}}{\delta\lambda}\right)\frac{\lambda_{0}}{D}.
\end{equation}
The MMT Shack-Hartmann WFS uses a $12\times12$ sub-aperture array
centered on the pupil, which means the maximum measurable unaliased
spatial frequency is 6 cycles$/D$, beyond which we cannot compute
the speckles. This corresponds to a fractional bandwidth of about
17\%, which is reasonably well matched to the L band filter. Well
inside the radial blurring limit, we are progressively more free to
ignore the effects of bandwidth and simply use monochromatic Fourier
optics \citep{1995ifo..book.....G} at the characteristic wavelength
$\lambda_{0}$. 

Depending on the brightness of the star compared to the background
noise and the shape of the speckle halo, the SNR of our measurement
varies widely across the focal plane. A long science camera exposure
will smooth the intensity fluctuations caused by the speckles beating
against the static halo, reducing their contrast and making them harder
to detect against the background noise and fluctuations. The WFS data,
converted to complex speckles, also has to be integrated to the science
camera's frame rate in order to compare the complex speckles with
the images. Our metric for signal is the ``visibility'' (Eq.~\ref{eq:visibility}),
but that alone does not express the fluctuating signal relative to
the background noise. The visibility is the phase-dependent intensity
variation relative to the mean intensity. The proper metric for comparing
the intensity variations, from which we derive all other information,
and the background noise is 
\begin{equation}
\mbox{SNR}=\frac{V\Phi_{I}}{\Phi_{noise}}=\frac{2\left|\left\langle \RR I\right\rangle \right|}{\Phi_{noise}\,\sqrt{\PHIsp}}.\label{eq:SNR}
\end{equation}
The average intensities, $\Phi_{noise}$ and $\PHIsp$, are unaffected
by the exposure time since they are sums of positive contributions.
The impact of exposure comes only from the $\left|\left\langle \RR I\right\rangle \right|$
factor, which we can analyze. Treating $I(t)$ and $\field(t)$ to
be instantaneous measurements, we write the effect of longer camera
exposures as an integral average of duration $T$ centered on $t=0$,
and use an ensemble average to estimate the effect on $\left|\left\langle \RR I\right\rangle \right|$.
In Eq.~\ref{eq:SNR} we replace $I$ with an integrated $I$ to represent
the time exposure 
\[
I(t)\rightarrow\frac{1}{T}\int_{-T/2}^{T/2}I(t+t')\dd t',
\]
and 
\[
\field(t)\rightarrow\frac{1}{T}\int_{-T/2}^{T/2}\field(t+t')\dd t'.
\]
Multiplying the down-sampled measurements and ensemble averaging,
the stabilized intensity becomes 
\[
\left\langle \RR(t)I(t)\right\rangle \rightarrow\frac{1}{T^{2}}\iint_{-T/2}^{T/2}\left\langle \psi(t+t_{1})I(t+t_{2})\right\rangle \dd t_{1}\dd t_{2}.
\]
Referring back to our simple model for the intensity, Eq.~\ref{eq:betterIModel},
we expand and multiply through by the field, performing the ensemble
average on each term. Based on our assumption that the speckle field
is ultimately a zero-mean Gaussian random process, the only term that
survives averaging is $I(t)=g\halo\field^{*}(t)$. Thus
\begin{equation}
\left\langle \RR(t)I(t)\right\rangle \rightarrow\frac{g\halo}{T^{2}}\iint_{-T/2}^{T/2}\MCF(t_{2}-t_{1})\dd t_{1}\dd t_{2},\label{eq:IntegratedStabTerm1}
\end{equation}
where $\MCF(\tau)$ is the complex speckle field's mutual coherence
function (MCF) 
\begin{equation}
\Gamma(t_{1},t_{2})=\left\langle \psi(t_{1})\psi^{\star}(t_{2})\right\rangle .\label{eq:MutualCoh}
\end{equation}
The MCF becomes time-invariant, depending only on the time difference,
$\Gamma(t_{2}-t_{1})$ with the symmetry $\MCF(\tau)=\MCF^{*}(-\tau)$,
so long as parameters such as wind, $r_{0}$, and AO performance remain
constant. We can normalize the performance by the assuming that the
camera exposures match the WFS, which gives the reference value of
$\left\langle \RR I\right\rangle $ as 
\begin{equation}
\left\langle \RR I\right\rangle \rightarrow g\halo\MCF(0)\equiv g\halo\PHIsp.\label{eq:ideal-stabIntensity}
\end{equation}
We divide Eq.~\ref{eq:IntegratedStabTerm1} by Eq.~\ref{eq:ideal-stabIntensity}
to estimate the effect of exposure time on SNR. Changing time integration
variables to sums and differences and using the MCF symmetry allows
us to write the effect of exposure on SNR as 

\begin{equation}
\mbox{SNR \ensuremath{\propto}}\frac{1}{T}\int_{0}^{T}\left(1-\frac{\tau}{T}\right)\Re\left\{ \frac{\MCF(\tau)}{\MCF(0)}\right\} \dd\tau.\label{eq:MCF-SNReffect}
\end{equation}
We will use this result with the actual MMT data in Sec.~\ref{sub:CSpeckles-from-WFS}. 

The residual speckle halo $\PHIsp(\thv)\equiv\Gamma(t_{1},t_{1};\thv)$
consists of a broad fitting error halo along with lag error speckle
plumes in the apparent projected direction of the wind. If there are
multiple wind streams along the line of sight, they will each contribute
their own speckle plume. Both fitting and lag error phase patterns
are carried across the pupil by their respective winds, with characteristic
effects expressed in the resulting speckles. For simplicity, we will
consider only one dominant wind stream with velocity projected onto
the pupil plane of $\wind$. Fitting error has minimal power in spatial
scales larger than $\ell_{AO}\sim D/\sqrt{N_{modes}}$ and the corresponding
halo is therefore dark within $\lambda/\ell_{AO}$, or at least relatively
constant, depending on details of the AO system. Processing lag error
contributes a residual wavefront that is proportional to the gradient
of the uncorrected wavefront dotted into the wind shift after the
lag (i.e. $\tlag\wind\cdot\nabla\phase_{0}$) since the wavefront
correction servo makes the same error in every iteration. This leads
to a plume of speckles in the direction of the wind that becomes brighter
as we look closer to the star. Depending on the details of the AO
system, the fitting error changes completely by the time the wind
has carried the wavefront by the AO correction scale $\ell_{AO}$.
Therefore, the fitting error speckles decorrelate after a timescale
of $\ell_{AO}/\left\Vert \wind\right\Vert \lesssim\tau_{fitting}\lesssim D/\left\Vert \wind\right\Vert $
(a fresh breeze of 20 m/s at the MMT would give timescales of 43 ms$\lesssim\tau_{fitting}\lesssim$325
ms). Lag error speckles are still affected by larger spatial scales
and can last much longer (App.~\ref{app:additive_speckles}). These
are certainly completely decorrelated by the time the wind has carried
the turbulent pattern by the outer scale, $\tau_{wind}\lessapprox L_{outer}/\left\Vert \wind\right\Vert $
(possible on the order of a second or more at the MMT). The outer
scale in all layers of the astronomical AO problem are often considered
to be less than 30~m, but may be longer in special cases. The decorrelation
timescales described here vary across the speckle halo around the
star, and only real data can tell the actual behavior. But we can
say a few things that ought to be robust statements. The fitting error
speckles should decorrelate fairly rapidly, while lag error speckles
should last significantly longer. 

The same is not true of the translation effect, which is systematic
and highly dependent on position. Translation of the residual wavefront
by an amount $\delta\xx$ causes any speckles at angular position
$\thv=\kkv/k$, $k=2\pi/\lambda$, to undergo a phase shift of $\delphase=-k\thv\cdot\delta\xx$.
(This ignores the overall change in coherence due to slightly different
areas of the wavefront being visible through the pupil at different
times, which is contained in the previously discussed portions.) Therefore,
a wind will cause a linear phase shift $\delphase=k\thv\cdot\wind\delta t$.
This affects all of the speckles together systematically, and shows
up in the MCF as 
\begin{equation}
\MCF(\tau)=\MCF_{0}(\tau)\e^{2ik\tau\thv\cdot\wind},\label{eq:MCF-spinningPhase}
\end{equation}
where $\tau=t_{2}-t_{1}$. 

We can now estimate the effect of science camera exposure time. We
will consider only a single wind stream of \textasciitilde{}10~m/s,
noting that for a moderate frame rate of $\sim30$ fps, a jet stream
contribution with a speed of \textasciitilde{}60~m/s will be highly
averaged over virtually all of the focal plane, appearing as incoherent
noise. For fitting error, the slow change in the MCF due to different
configurations being present within the pupil has a timescale of \textasciitilde{}650~ms
but likely much shorter due to the AO iterations and other effects.
Lag error speckles within the control radius but outside of the PSF
core arise from spatial scales smaller than the pupil diameter $D$
and are most likely uncorrelated much beyond the pupil. Therefore
they too have a maximum lifetime of hundreds of milliseconds, which
is common. The systematic phase rotation caused by the wind increases
most rapidly in the direction of the speckle plume and is not seriously
detrimental until the phase wrapping from the time center of the exposure
to the endpoints is $\pi/2$, beyond which negative contributions
are included. This means that the phase rotation timescale, providing
us with a conservative exposure time limit, is when $\pi>k\left|\thv\cdot\wind\right|\tsci$
or 
\begin{equation}
\tsci<\frac{D}{2\left|\frac{\thv}{\lambda/D}\cdot\wind\right|}.\label{eq:ExpLimit-Wind}
\end{equation}
Using the WFS Nyquist radius for the MMT $12\times12$ WFS along either
axis, $\left\Vert \thv\right\Vert =6\lambda/D$, and a wind of 10
m/s, we find that our exposure limit is about 54~ms. Assuming continuous
exposures, that corresponds to about 18.5 fps. Referring back to the
effect of decorrelation on the SNR, Eq.~\ref{eq:MCF-SNReffect},
we can see that our limit only drops the SNR by 64\% at the limiting
radius. Using a higher frame rate camera is therefore not extremely
important for sensitivity, but would be for an increased measurement
radius or for higher wind speeds.

\section{Observations and Analysis}

\label{sec:observations}To measure the complex halo at the science
camera, we require three capabilities: (1) acquire short exposures
with the science camera, (2) AO wavefront sensor data, and (3) provide
a mechanism for synchronizing the resulting data sets together. Our
mid-IR camera is capable of reading out small regions of the sensor
at frame rates in excess of 30 Hz, and our high-speed AO WFS and subsequent
processing software has an engineering diagnostic mode capable of
saving the full system telemetry, including raw WFS pixels and computed
slopes. Tight synchronization between the science camera and the AO
system is not provided by the MMT and is added using system handshaking
and logging modifications. We describe our solutions to these generic
problems here, as well as other difficulties to do with the engineering
state of the MMT at the time of our observations.

\subsection{Facilities}

\subsubsection{The MMT AO System}

\begin{figure}
\centering{}\includegraphics[width=1\columnwidth]{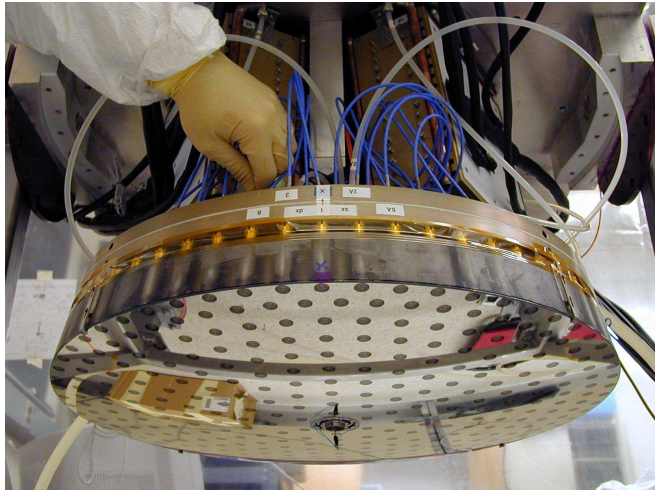}\caption{\label{fig:MMT-DM}The MMT's deformable secondary mirror with an oblique
view of the 336 voice coil actuators and their modified hexapolar
placement. }
\end{figure}

The MMT Adaptive Optics system \citep{Wildi02} is the world's first
telescope to use a deformable secondary mirror to provide wavefront
correction. This approach minimizes the number of warm optical surfaces
between the sky and the science camera, greatly reducing the thermal
emissivity of the telescope, and makes the 6.5~m MMT aperture competitive
with larger telescopes for thermal infrared observations \citep{2000SPIE.4007..167L}.
The 640~mm diameter deformable secondary consists of a thin shell
mirror 2~mm thick, supported above a Zerodur reference body by a
fixed central hub, and deformed by 336 actuators in a modified hexapolar
pattern (Fig.~\ref{fig:MMT-DM}).
 The actuators provide non-contact forces to the shell via electromagnetic
voice coils acting on magnets attached to its inner surface. The gap
between the shell and the reference body is measured by capacitive
sensors at each actuator, and is actively maintained by a 40 kHz servo
in a dedicated mirror controller. The typical time for the shell to
reach a desired position is less than 1~ms.

The MMT AO processing is performed by a real-time Linux computer system
\citep{Vaitheeswaran08} which reads the WFS camera, computes the
required wavefront updates, and sends updated commands to the deformable
secondary controller, all synchronously clocked by the WFS frame rate.
The 12$\times$12 subaperture Shack-Hartmann WFS \citep{Mcguire99}
is normally operated at 527 Hz. The AO computer calculates the wavefront
slopes, reconstructs an estimate of the wavefront, applies a conservative
modal filter, and sends the updated actuator commands to the deformable
secondary controller. 

The MMT's Shack-Hartmann WFS uses a $12\times12$ lenslet array to
image the starlight within each subaperture onto the center of a binned
$2\times2$ pixel ``quad-cell.'' Local wavefront slopes cause the
image to shift, changing the relative amount of starlight entering
the various quad-cell pixels. The pixels are exposed, read out, summed,
differenced, and mapped through a lookup table matched to the seeing
level, yielding estimates of the $x$ and $y$ wavefront slopes over
each of the 144 subapertures \citep{Hardy98}. The pixel sums and
differences are normalized by the sum of the quad-cell counts, plus
a small bias term. The bias ensures that unilluminated quadcells generate
zero slope estimates. The resulting 288 $x$ and $y$ slopes are serialized
into a single column vector (the ``slopes vector'' $\SLOPES$).
The AO system estimates the residual wavefront by multiplying the
slopes vector with a wavefront reconstructor matrix, $\RECONao$,
resulting in an estimate of the residual wavefront error at the actuator
positions: $\delta\DISPLACEMENTS=\RECONao\SLOPES$. An AO reconstructor
typically has a combined legacy of optical measurements and analytical
processing, characterized ultimately by a number of singular value
decomposition (SVD) modes \citep{Brusa03a}. The MMT AO reconstructor
uses lowest-energy mechanical modes of the thin shell as a basis set.
This restriction means that the wavefront spatial frequencies are
not uniformly corrected within the modal control radius of the mirror,
implemented as a safety precaution. Subsequent developments with the
LBT AO system \citep{Esposito10} and Magellan observatories are less
constrained. The wavefront correction is post-processed using a ``modal
filter'' to redundantly ensure that damaging stresses are not applied
to the shell. Once calculated, the residual wavefront estimates are
multiplied by an overall (scalar) gain factor and added to the correction
already applied to the mirror. The result is then sent to the mirror
controller where it is applied with its own high-speed servo control
loop. 

\begin{figure}
\begin{centering}
\includegraphics[width=6cm]{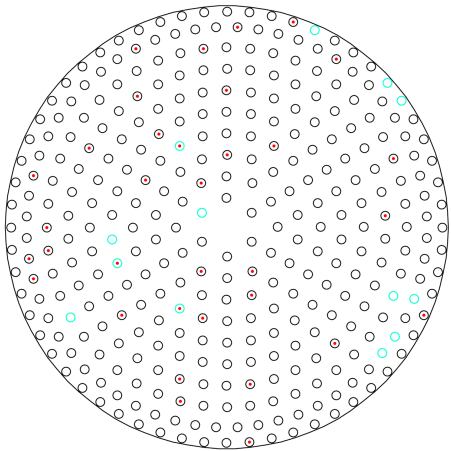}
\par\end{centering}

\caption{\label{fig:MMT-actuators}Locations of the MMT deformable secondary
mirror actuators (circles). At the time of our measurements, there
were 13 disabled actuators (light gray circles) that float to a point
between the adjacent actuators. 34 actuators had problematic or failed
capacitive sensors (dots), but were still able to be driven in open
loop by the mirror controller. For the 56-mode correction applied
by the AO control servo, having 10\% failed actuators has only a marginal
effect. }
\end{figure}

There were several issues with the MMT AO system at the time of observation.
The gains on individual mirror actuators had not been recalibrated
for several years and in some cases had drifted by as much as tens
of percent. Thirteen out of the 336 actuators were non-functional
(Fig.~\ref{fig:MMT-actuators}).
 The loss of these actuators did not significantly impact low-order
correction modes as the DM shell above a deactivated actuator floats
to a position interpolating that of its neighbors. Another 34 actuators
had inoperative capacitive sensors, requiring their positions to be
set in the high-speed mirror controller by \emph{dead reckoning} rather
than using the closed-loop control servo. In addition to some mount
vibrations, our observations include significant pointing swings deliberately
introduced as a signal for calibration of a pyramid wavefront sensor.
These vibrations have amplitudes in excess of 100 mas (i.e. on the
order of a science camera image diffraction width).

\subsubsection{The Clio mid-infrared Science Camera}

Our science camera was the mid-infrared \emph{Clio} system \citep{2004SPIE.5492.1561F,2006SPIE.6269E..27S}.
Light from the deformable secondary directly enters Clio's dewar through
a tilted dichroic window, forming an $f/15$ image at the first focal
plane. A re-imaging lens forms another pupil plane, usually for bandpass
filters and Lyot stops, but also where Clio's Apodizing Phase Plate
(APP) is located \citep{CLIO_ApJ_Kenworthy_etal_2007}. A final lens
images the pupil plane onto the focal plane imaging sensor. Clio's
sensor is a HAWAII-1 HgCdTe array with 18.5 $\mu$m square pixels
cooled to 75.6~K. The detector gain is 4.9~$\e^{-}$/dn with a bias
of 3700~dn and a saturation level of 55,000~dn, giving a full-well
capacity of 51,000~dn (250,000~$\e^{-}$). The read noise for a
single frame is 19~dn (93~$\e^{-}$). The dark current is 50~dn/s
(245~$\e^{-}$/s) at 75.9~K. Clio's plate scale was measured as
0.0299''/pixel or 4.0 pixels$/\left(\ld\right)$ in L band. Clio
is controlled by a computer running Linux, taking exposures and saving
FITS image cubes asynchronously from the AO system. Data is taken
using a special 54$\times$108 ``sub-stamp mode'' to achieve a higher
frame rate.

\subsection{Synchronizing WFS and Science Data}

Since the speckle phases change rapidly, both randomly and in a deterministic
way depending on position within the field and the projected speed
and direction of the wind, small synchronization errors can cause
both random and systematic biases in the measured halo phase. Without
taking particular care, we might expect the phase error to be determined
by how much the speckle phase changes during the science camera's
exposure time. However, this can be reduced by carefully synchronizing
the WFS frames to the start and end of the science frames, thereby
achieving phase accuracies more in line with the the speckle phase
change during a single WFS frame. This improvement (a factor of 6.5
in this dataset) only helps reduce biases. The loss of coherence caused
by the science camera exposure time is not recoverable and still reduces
the measurement SNR. However, by removing systematic effects, averaging
will still yield more accurate measurements at the higher frame rate. 

All AO system updates occur at the WFS frame rate. In an engineering
diagnostic mode, the AO host computer buffers 10,000 frames of engineering
data into a set of RAM-based circular buffers. Once these buffers
are filled, a separate processing thread writes them out to the AO
computer's hard disk. The MMT WFS frames are normally taken at a steady
rate, but an engineering issue resulted in every other WFS frame being
dropped from the data telemetry, resulting in the system operating
at half its normal speed.The result was that the normally 527 frames/s
rate was reduced to approximately 263.5 frames/s with a slopes file
saved every 38~s. To avoid timing problems and other possible buffer
overrun issues, we did not include any science data sets that ran
across WFS slopes files. These issues do not impact our technique.
Clio buffers 100 contiguous exposures in RAM before saving them to
disk. When the RAM-based image buffer is filled, the camera acquisition
halts while the images are saved. The non-standard small images were
acquired at approximately 40.5 frames/s, filling the 100-image buffer
in about 2.5~s and saved to disk every 3~s. Within each Clio data
cube, the images are continuous sequential exposures, with each exposure
ending as the next began. 

Both the AO and Clio control computers run Network Time Protocol (ntp)
clients. The resulting filesystem and internal header timestamps are
directly compared between computers, allowing us to uniquely associate
log entries and data files. Using ntp alone does not guarantee synchronization
to the WFS frame level. The AO engineering diagnostics files contain
only data and no internal timestamps or other useful FITS header information.
The primary identifying information was a unique file name which included
a numerical code derived from internal counters and the time of the
first frame, recorded to 1 second accuracy. The operating system also
encoded timestamps in the filesystem inodes when the file was written.
Since these are volatile upon copying and archiving, we saved them
to a file after the run using the Linux command ``\texttt{ls -lt
-{}-full-time > timestamps.log}'' which preserved the file creation
times in ISO format to the accuracy configured in the kernel (only
1-second accuracy for both systems). The Clio file creation timestamps
are also saved as a fallback procedure, but the primary record there
was a timestamp saved in the FITS file header. 
\begin{figure}[h]
\begin{centering}
\includegraphics[width=0.9\columnwidth]{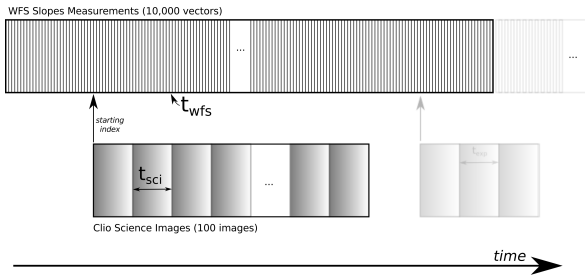}
\par\end{centering}

\caption{\label{fig:Synchronization-of-the}WFS and science data cubes have
asynchronous placement and non-guaranteed timing. Although clock synchronization,
network handshaking, timestamps, and logging procedures were implemented,
a data-based cross correlation procedure was also required to ensure
proper synchronization. }
\end{figure}

In addition to the timestamps, we implemented a simple network handshaking
protocol to record the WFS frame information to higher accuracy. At
the beginning of a set of science images, The science camera control
computer sends a UDP packet to the AO host computer. Once received
by the AO system, this information is written to the AO log file along
with the base name of the WFS engineering files currently being recorded
and the current WFS frame number. This information is later extracted
from the log and used to align the data sets to within a few WFS frames.
The handshake procedure contains an unknown delay from the time Clio
transmits the UDP packet and the AO system wrote the current WFS frame
offset to the log file. This correlation gave us the mean number of
WFS frames per science exposure, $\left\langle \tsci/\twfs\right\rangle $,
and the lag between the logged frame number and the correct value. 

Using the images from a single 100-frame Clio file, we determined
the location of either the star's peak pixel or its centroid. We then
interpolated this time series based on an assumed value for $\left\langle \tsci/\twfs\right\rangle $
to the WFS frame rate, allowing a cross correlation with the mean
$\theta_{x}$ and $\theta_{y}$ time series determined from the appropriate
subset of the WFS slope file, extended before and after by a generous
set of additional frames. The concurrent engineering tests were introducing
a lot of tip-tilt noise, giving us a strong signal for the timing
calibration. Our exposures were short enough that this did not cause
us any problems. Even so, the normal tip-tilt noise in the AO system
would have been adequate for this determination. The science camera
and the WFS are not necessarily aligned, but were in this case. Even
so, we correlated all four combinations between the two inputs to
ensure that we had the coordinate mapping correct. By varying the
presumed ratio of frame rates, we determined $\left\langle \tsci/\twfs\right\rangle \approx6.48\pm0.03$.
Thus, in 100 continuous science exposures, assuming the center exposure
was correctly placed, the start and end exposures have a placement
uncertainty of $\pm1.5$ WFS frames or 5.7 ms. The cross correlation
analysis also determined the lag from the handshake protocol to be
3 WFS frames. Once calibrated, the handshake protocol alone was sufficient
to synchronize the Clio images with the WFS data to within a WFS frame
on the average. Without the random WFS readout timing problems encountered
during our test run, and a measured frame rate for the science camera,
all the exposures would be placed relative to the WFS with equal precision,
and overall accuracy would be determined by the accuracy of the calibrated
handshake. This is expected to be within a single WFS frame, which
is the limit of accuracy for the system. As we shall see below, the
speckle coherence time varies across the field, but the minimum values
are comparable to the timing accuracy errors at the bounds of the
exposures. The result is that there were some coherence losses in
SNR due to temporal misalignment between the datasets, but they were
not serious enough to limit our results.

\subsection{Science Camera PSFs}

\begin{figure}
\centering{}\includegraphics[width=1\columnwidth]{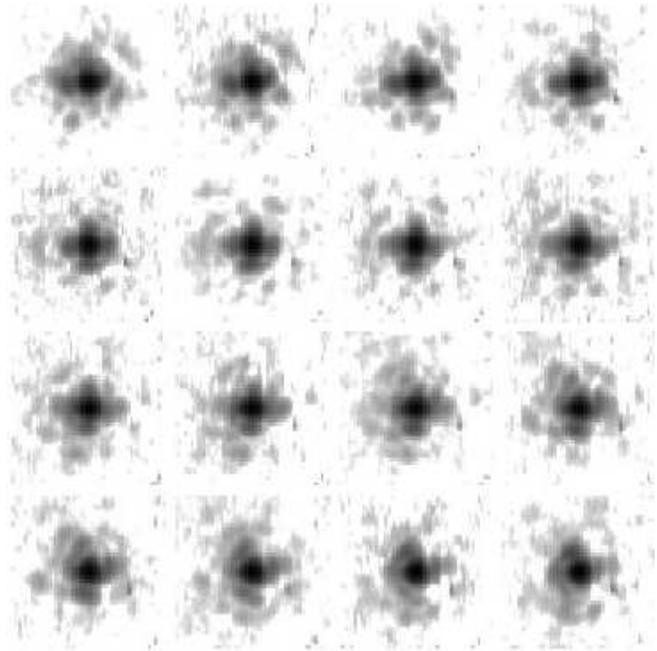}\caption{\label{fig:Example-Clio-images}Clio star images. These 16 sequential
images are shown with a 3 decade log grayscale.}
\end{figure}

The observed star is a 7.4 mag G5 star at zenithal angles ranging
from 13 to 14 degrees (Fig.~\ref{fig:Example-Clio-images}). 
Clio was configured for direct imaging with a Barr MKO L' band filter
\citep{Tokunaga02}, centered at 3.8~$\mu$m and a bandpass of 0.7~$\mu$m.
At the center wavelength, the Clio pixels subtend $\lambda/4.0D$.
The $54\times54$ pixel region readout from the sensor was calibrated
with dark and flat images, with noisy or bad pixels flagged and replaced
by the median of a $3\times3$ window centered on the pixel. Image
vibrations were removed by shifting the PSF center with bilinear interpolation
to a fixed location. The PSF peak was located using correlation with
a Gaussian reference peak. The pointing vibration during our observations
had a swing of approximately 1.5 -- 2 $\ld$ which contributed only
a small amount of image motion blur during any given Clio exposure.
This vibration did not significantly impact the data reduction process,
but did have the effect of dithering over any bad pixels, reducing
their influence. It was not possible to precisely know at observation
time where we were in the WFS buffer, so no attempt was made to avoid
the times bridging slopes files. To avoid any possible timing complexities,
any Clio image cubes that bridged WFS file boundaries were simply
discarded.

\subsection{Complex Speckles from WFS slopes\label{sub:CSpeckles-from-WFS}}

We first use the WFS measurements to estimate the varying wavefront
displacement in the pupil plane $z(\xx,t)$, where $\xx$ is the position
in the pupil plane. Using the science camera filter band's mean wavelength
$\lambda$, we compute the phase shift caused by the displacement,
$\pphase(\xx,t)=kz(\xx,t)$, where $k=2\pi/\lambda$. Ignoring scintillation,
we write a unit amplitude complex field in the pupil plane as $\exp\left\{ i\pphase(\xx,t)\right\} $,
pass it through the pupil stop $\pupil(\xx)$, and then use Fourier
optics \citep{1995ifo..book.....G} to find the complex halo in the
image plane 
\begin{equation}
\halo(\kkv,t)=\int\e^{i\kkv\cdot\xx}\e^{i\pphase(\xx,t)}\pupil(\xx)\,\dd^{2}x\label{eq:Fourier-Optics}
\end{equation}
where $\kk$ is spatial frequency in the pupil plane and $\thv=\kkv/k$
is the corresponding angular coordinate (in radians) measured from
the star. Over a given observation period, we can make a distinction
between the ``static halo'' and ``speckles'' as 
\begin{equation}
\halo(\kkv,t)=\underbrace{\overline{\halo}(\kkv)}_{static\, halo}+\underbrace{\psi(\kkv,t)}_{"speckles"}.\label{eq:Halo-dissection}
\end{equation}
(A more detailed discussion of the separation of the halo into its
static and speckled components is given in App.~\ref{app:additive_speckles}.)
Basic Fourier optics is a very simple model for an imaging system,
with no place to properly introduce aberrations that occur after the
starlight passes the pupil without significantly increasing the complexity
of the optical wave model. Aberrations that occur anywhere other than
the pupil plane (or images thereof), will affect the halo and speckles
differently depending on where the star appears in the camera's field
of view. Since the field of interest in high-contrast imaging is very
small, we only introduce negligible errors by treating all aberrations
as having taken place in the entrance pupil plane, including any downstream
aberrations not seen by the AO system (i.e.~NCP aberrations). It
is because of this that we may measure the complex halo in the focal
plane and then use Fourier optics to find an equivalent distorted
field in the pupil plane. Since we do not know the NCP aberrations
\emph{a priori}, we must compute the speckle halo without them. Depending
on the magnitude and nature of the NCP errors, this can have a large
effect, distorting both the static and speckle halos. At higher Strehl
ratios (where the peak of the diffraction-limited PSF is much brighter
than the speckles), each individual complex speckle can be thought
of as a translated, scaled, and phase-shifted copy of the speckle-free
halo. Therefore, including the NCP aberrations would mostly just alter
the ``speckles on the speckles'', adding noise to the speckle halo,
but having only a minor effect on the brighter speckle peaks. The
NCP errors also alter the static halo, often in ways which cannot
be ignored. However, by estimating and removing the idealized static
halo from our calculation, we are left with just the speckles. This
justifies why we can ignore the NCP aberrations when computing the
speckles, allowing us to bootstrap our interferometric measurement
of the true static halo. If we need more accuracy, we can iterate
the process by using the first estimate of the NCP aberrations in
a second calculation of the complex speckle halo.

As we collect WFS data, we repeatedly carry out the above steps to
build a ``data cube'' of complex halos that do not include the effects
of NCP errors, but do include residual AO speckles along with a simplified
static halo. This static halo is presumed common to all frames of
the data cube ($\halo_{nmw}\equiv\halo(k\thv_{nm},t_{\mathit{w}})$,
where $\thv_{nm}$ is the grid of pixel coordinates) and can be estimated
by averaging over the time index. Since we force the time average
of the speckles to be zero over each data cube, we are actually pushing
any estimation error onto the derived static halo. This is a consequence
of processing with smaller data cubes, and the resulting individual
static halo estimates from multiple data cubes, when averaged, should
lead to a more accurate answer. 

As mentioned earlier, the WFS slopes are related to wavefront displacements
by a ``reconstructor matrix.'' This is essentially an integrator,
working on the set of $x$ and $y$ slopes returned from the WFS sub
apertures. The $x$ and $y$ slopes from the $12\times12$ WFS are
serialized into a single $288\times1$ ``slopes vector'', $\SLOPES$,
which is integrated into a wavefront by multiplying by a ``reconstructor
matrix'' $\RECON$: $\DISPLACEMENTS=\RECON\SLOPES$. Since the MMT
AO wavefront reconstructor, $\RECONao$, only corrects a limited number
of modes (currently 56), it is not sufficient for our complex halo
calculation. If we consider each pair of modes to be the equivalent
of controlling the real and imaginary parts of the complex amplitude
of a speckle, each of which has a width of roughly $\ld$, then 56
modes will take us out to roughly $3.7\ld$ from the star, while the
WFS is capable of at least $6\ld$. To allow us to properly recover
the residual wavefront to the accuracy of the WFS rather than being
limited by the AO system, we require a better reconstructor, $\RECONspeckles$.
Since our reconstructor is not used to drive the DM, nor is it intended
to be iterated by being placed inside of an AO servo loop, it is not
as important to be conservative about issues such as modal gain, making
it easier to derive an adequate reconstructor. Also, since the computed
displacements are not going to be used to drive the actual DM, we
can choose to reconstruct the wavefront at more conveniently-placed
locations across the pupil: e.g. on a square grid, instead of the
actual hexapolar locations of the physical DM. But here, to facilitate
direct comparison with the AO system's data as well as other live
reconstructor tests, we used the physical actuator locations. For
each WFS slopes vector, the estimated residual wavefront displacements
at the actuators are given by $\DISPLACEMENTS=\RECONspeckles\SLOPES$. 

We estimated and removed the mean $x$ and $y$ slopes from the slopes
vector before multiplying by $\RECONspeckles$, resulting in a vector
of wavefront displacements with tip-tilt removed. For computing the
complex pupil field and halo, we interpolated the wavefront displacement
estimates at the actuators to a 4 cm square grid using Delauney triangularization
and cubic interpolation, yielding a wavefront displacement $z_{nm}$
at $\xx_{nm}=(n,m)\delta x+\xx_{00}$. The complex pupil field was
computed by $\pfield_{nm}=\pupil_{nm}\exp\left(2\pi iz_{nm}/\lambdasci\right)$,
where $\pupil_{nm}$ is the pupil transmission mask interpolated to
the grid coordinates. The resulting complex mesh was zero-padded to
$N\times N$ and Fourier transformed, giving focal plane samples spaced
by $\delta\kappa=2\pi/N\delta x$ and angular spacing of $\delta\theta=\lambdasci/N\delta x$.
We adjusted $N$ to match the L-band Clio plate scale of $0.25\ld$.
The focal plane field was computed using 2-D FFTs and the results
kept in a complex data cube $\halo_{nmw}\equiv\halo(\kappa_{x},\kappa_{y},t_{\mathit{wfs}})$.
This processing was performed for each set of 100 science camera images
($\sim2.5$~s), with the corresponding set of WFS frames selected
and used to compute a halo data cube spanning the science camera exposures
at the WFS frame rate. Since the synchronization between the WFS data
and the science camera is only accurate to about one frame from the
center of a 100-exposure image set, but the duration of the individual
exposures is 6.5 frames, the starting frame index is assumed to be
on a frame boundary, with subsequent exposures placed on the timeline
as they fell. The full-speed complex halo was then down-sampled to
the science camera frame rate by summing the complex frames, including
linearly-weighted end frames according to the computed endpoints.
Finally, since the actual piston and tip-tilt may have changed during
a single science camera exposure, we performed the complex halo sums
before normalizing to the peak phase. Since tip-tilt was removed in
the slopes before computing the halo, motion blur was not fully represented,
but the image wander is much smaller than $\ld$ during any single
science camera exposure and the error introduced is small. The complex
halo estimate at the science frame rate was 
\begin{equation}
\widehat{\halo}_{nm\mu}=\e^{-ip_{\mu}}\sum_{w\in\tsci(\mu)}\halo_{nmw}\label{eq:downsampledHalo}
\end{equation}
where $p_{\mu}=\left.\arg\left\{ \sum_{w\in\tsci(\mu)}\halo_{nmw}\right\} \right|_{core}$.
Since $\left\langle \phi\right\rangle =0\nRightarrow\arg\left\{ \left\langle \e^{i\phi}\right\rangle \right\} =0$,
core phase referencing was performed on the complex WFS PSF halo.

Since the individual complex halo cubes were so short, we did not
force the mean speckle field to be zero for each individual cube,
giving a changing estimate for the simplified static field used in
the calculation. Instead, we used an estimate of the ensemble average
of the halo, (i.e. $\SR^{1/2}\Psi_{0}$ as described in App.~\ref{app:additive_speckles})
and used it to estimate the speckle halo over each cube. This is better
than using the per-cube average since the estimation error appears
as a constant across the pupil rather than randomly textured (Sec.~\ref{sub:Pupil-Field-Est-with-NCP}).
Our resulting speckle field was  
\begin{equation}
\psi_{nm\mu}=\widehat{\halo}_{nm\mu}-\SR^{1/2}\Psi_{0}.\label{eq:computedSpeckleField}
\end{equation}
Note that while this is no longer zero mean on the scale of individual
image cubes, it should approach zero mean in the ergodic limit of
many data cubes. The speckle field is now able to be directly compared
with the science camera image cubes as they are synchronized and have
the same sampling in space and time. 

\begin{figure}
\centering{}\includegraphics[width=1\columnwidth]{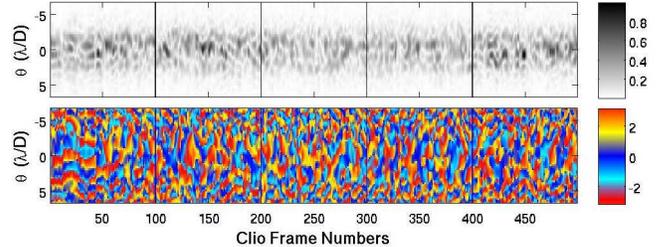}\caption{\label{fig:Speckle-Cut}A cut through the computed speckle field for
five non-contiguous Clio image sets. The upper panel shows the speckle
amplitudes as a function of time while the lower panel is the speckle
phase relative to the light in the PSF core. The 100 Clio frames of
each set are contiguous, but not across boundaries at multiples of
100. The speckle amplitudes are arbitrarily normalized to the brightest
value.}
\end{figure}

\begin{figure}
\begin{centering}
\includegraphics[width=5cm]{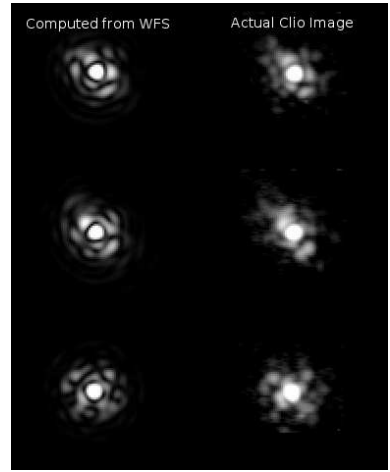}
\par\end{centering}

\caption{\label{figComparison-WFS-Clio}Star images computed from the residual
WFS measurements (left column) alongside the corresponding Clio images
(right). The computed images mainly differ from the actual images
in optical bandwidth and non-common-path (NCP) aberrations. The computed
images are monochromatic and assume that the estimated wavefront seen
by the WFS is the only source of aberration. }
\end{figure}

Before continuing on to compute the static halo, it is useful to look
at the behavior of the complex speckles and their spatial and temporal
statistics to understand how our result is affected by exposure time,
wind, etc. Fig.~\ref{fig:Speckle-Cut} shows an angle-time cut through
the center of the speckle field in the projected direction of the
wind to illustrate the lifespan of the speckles, with both amplitude
and phase variations.
 The speckle amplitude variation timescale is reasonably consistent
across the speckle cloud, while the speckle phase has a similar pattern
superimposed on a steady phase rate that increases as we move away
from the star. These speckles coherently add to the as-yet unknown
static halo, the intensity being recorded by the science camera. We
can make a visual comparison of the science images frame-by-frame
to the WFS-based images using $\left|\widehat{\halo}_{nm\mu}\right|^{2}$,
which is based on our simplified no-NCP error model (Fig.~\ref{figComparison-WFS-Clio}).
This figure illustrates if we are indeed computing speckles when and
where they were actually observed, and the extent of any deviations.
For each row in Fig.~\ref{figComparison-WFS-Clio}, the PSF derived
from the WFS slopes is shown on the left, while the corresponding
Clio image is on the right. The two images exhibit similar behavior,
but with the single-wavelength computed PSF having more well-defined
speckles than the actual science images due to the bandwidth of the
science images. The speckles in the computed images may also be more
or less prominent than the real images since the reconstructor gain
may be incorrect as a function of position around the star (i.e.~a
possibly incorrect modal gain). Also, since NCP aberrations and low-order
effects such as defocus are not included in the computed PSFs, there
might be noticeable differences between computed and real images,
as well as complicated differences with the speckles. In our present
case the comparison is reasonably good, providing a sanity check that
the synchronization and WFS-to-speckle calculation was performed correctly.

The various statistics are straightforward to compute from the image
and complex speckle data cubes. The intensity variance is 

\begin{equation}
\sigma_{I}^{2}=\frac{1}{N}\sum_{w=1}^{N}\left(I_{nmw}-\frac{1}{N}\sum_{w'=1}^{N}I_{nmw'}\right)^{2};
\end{equation}
the mean speckle halo intensity is
\begin{equation}
\PHIsp=\frac{1}{N}\sum_{w=1}^{N}\left|\field_{nmw}\right|^{2};
\end{equation}
the variance in the speckle halo's intensity 
\begin{equation}
\VARsp=\frac{1}{N}\sum_{w=1}^{N}\left(\left|\field_{nmw}\right|^{2}-\Phi_{speckles,nm}\right)^{2};
\end{equation}
and the stabilized intensity is 
\begin{equation}
\left\langle \RR I\right\rangle =\frac{1}{N}\sum_{w=1}^{N}\field_{nmw}I_{nmw}.
\end{equation}
The scale factor $\gain$ is computed using Eq.~\ref{eq:scale-factor}
and the static halo $\halo$ is given by Eq.~\ref{eq:TheAnswer}.

\begin{figure}
\begin{centering}
\includegraphics[width=1\columnwidth]{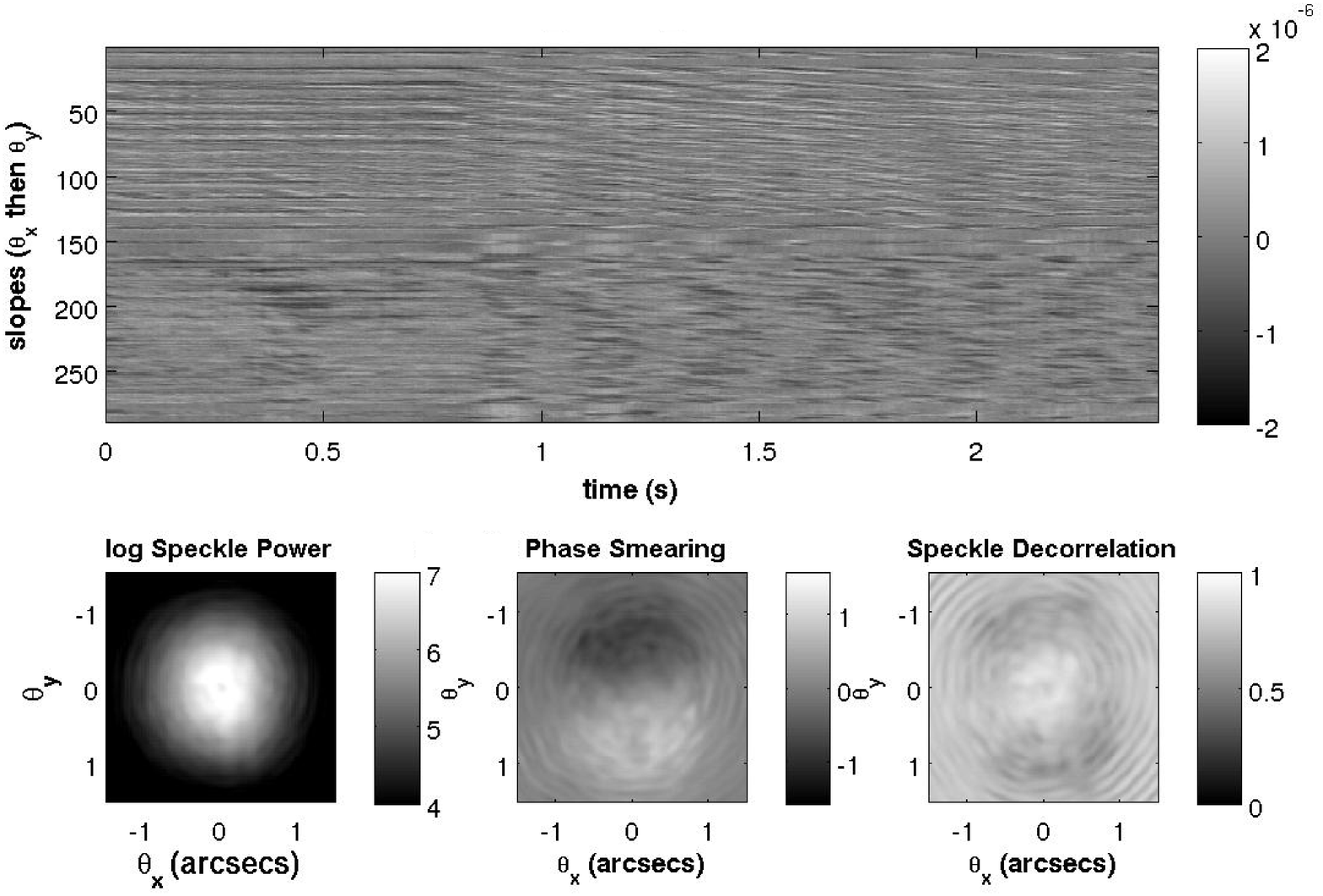}
\par\end{centering}

\centering{}\includegraphics[width=1\columnwidth]{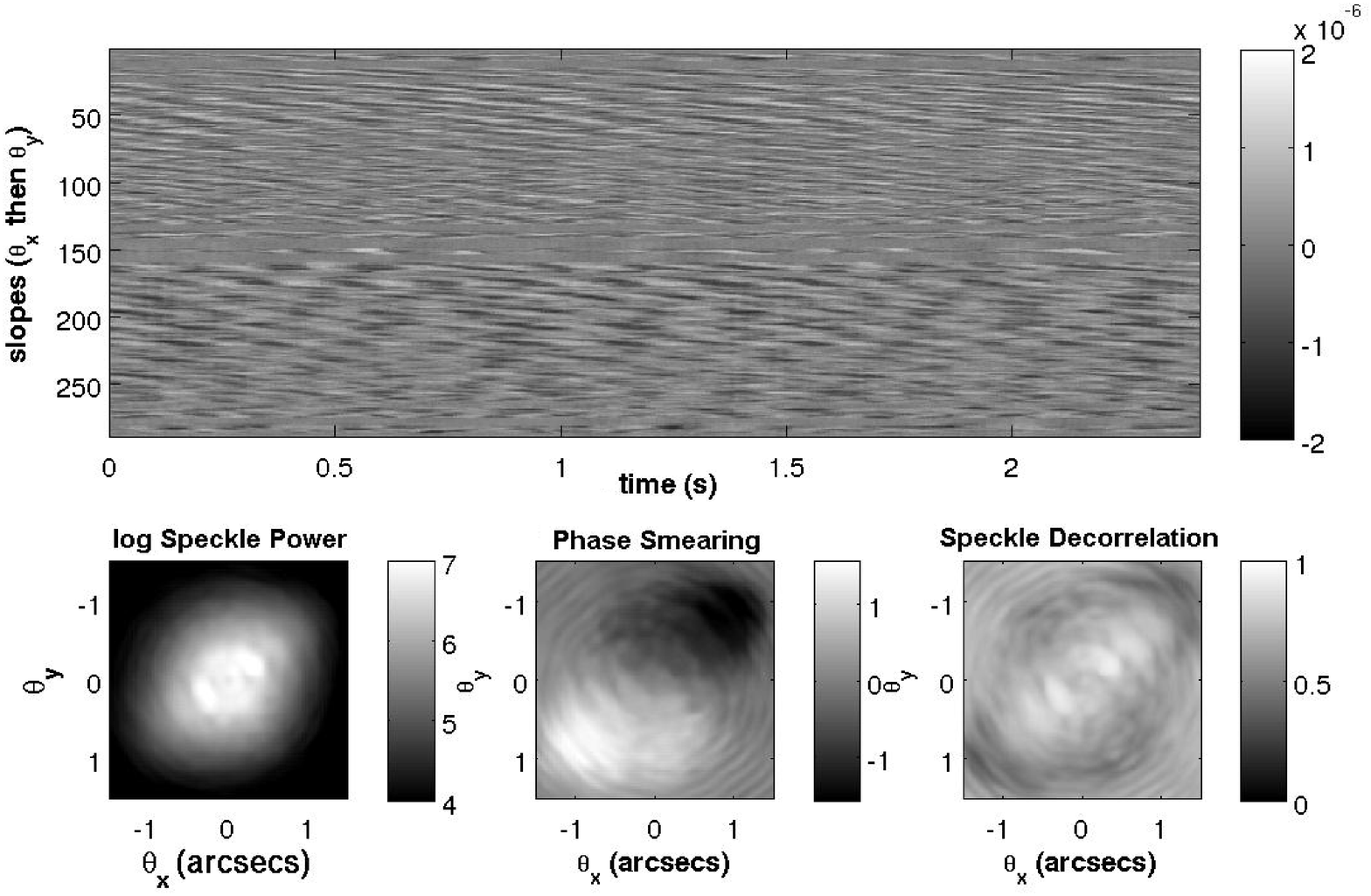}\caption{\label{fig:SlowFastWindExample}The effect of wind speed on the distribution
and evolution of the speckles. Two 2.5~s WFS data sets are shown,
with the 2.5~s of WFS slopes vectors organized as a $288\times650$
matrix. The slopes were reconstructed into residual wavefronts and
used to compute the complex speckle halo. In the top set of images
the wind is slower, while in the lower set the wind is gusting to
more than twice the upper speed. The speckle halo's mean intensity
is shown on the left. The slower wind case is dominated by the relatively
isotropic fitting error, while the faster wind shows a prominent lag
error speckle plume in the (shifted) direction of the wind. The average
speckle intensity is shown using a square root scale. The systematic
phase smearing during a \emph{Clio} exposure is is shown in the center
image. Finally, the speckle decorrelation on the right shows the loss
of speckle coherence at a lag roughly equal a Clio exposure. }
\end{figure}

The wind speed gusted by more than a factor of 2 and changed direction
by more than 45 degrees during the observation, causing clearly-visible
effects in the speckle field statistics. When the wind was slower
(say 5--10 m/s), the speckle cloud was more isotropic with a form
characteristic of fitting error, while stronger wind gusts enhanced
the plume of lag-error speckles. Fig.~\ref{fig:SlowFastWindExample}
shows two selected times with slower and faster wind. For each case
the figure shows the WFS slopes and three derived speckle statistics.
The $12\times12$ sub-aperture Shack-Hartmann WFS slopes were organized
into 288-element vectors and concatenated into a matrix of slopes
(upper images). The speckle halo was computed for each slopes vector
and the three statistical images were averaged from the 650 complex
speckle frames. The lower left images are the average speckle power
centered on the star, showing the morphological change between the
more isotropic fitting error halo and the lag error speckle plume.
The middle image is the average amount of phase rotation seen during
a Clio exposure, computed from the two-time MCF (Eq.~\ref{eq:MutualCoh}):
$\delphase(\thv,\tsci)=\arg\left\{ \Gamma(\thv,\tsci)\right\} $.
The phase rotation is mostly a systematic translation effect, while
the loss of coherence due to random effects is better described by
the drop in the magnitude of the MCF $\left|\Gamma(\thv,\tau)/\Gamma(\thv,0)\right|$,
shown in the lower right images in Fig.~\ref{fig:SlowFastWindExample}
for a lag of $6\,\twfs$. The linear grayscale runs from completely
incoherent (black) to completely coherent (white). The results show
that while fitting error speckles rapidly lose coherence, lag error
speckles remain coherent longer in the direction of the wind. This
is due to the well-known phenomenon of the AO system repeatedly making
the same lag-induced correction error as the aberrations are carried
across the pupil. 

\begin{figure}
\centering{}\includegraphics[width=1\columnwidth]{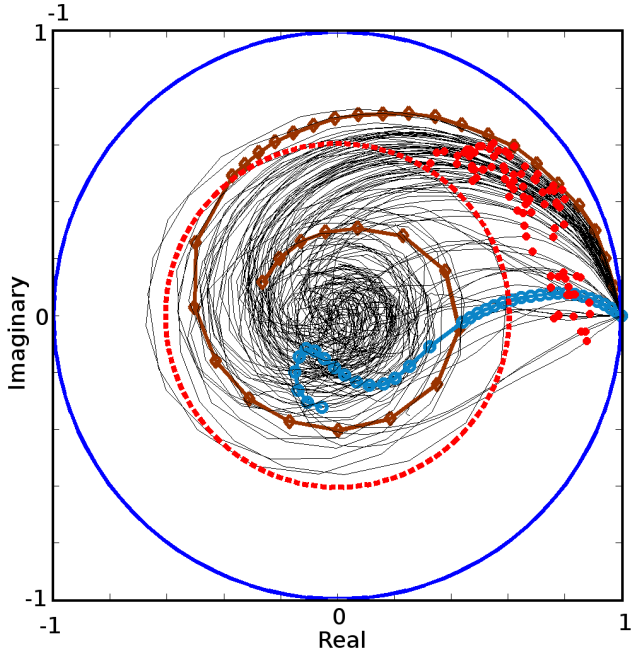}\caption{\label{fig:Speckles-MCF-4panel1}Normalized mutual coherence function
$\MCF(\tau)/\MCF(0)$ for a single point in the second Airy ring.
Two example cases are highlighted, one where the angle between the
wind vector and the pixel vector is small, (brown with diamonds) and
one where the angle is significantly larger (blue with circles). For
each of the WFS-derived speckle data cubes, $\MCF(\tau)$ was computed
at lags $\tau:\{0,1,2,\cdots,20,25,30,\cdots,100\}\twfs$. The speckles
always decorrelate with time, but wind causes a systematic phase shift
in a direction related to the projection of the wind on the selected
point in the focal plane. The interesting cases where the angle swerves
from one sign to the other are caused by more than one wind stream
competing. The values of the MCF at the end of the Clio exposure are
shown as filled red circles, showing that for this pixel the phase
smearing was never a serious issue. The speckle coherence time, $\tnaught$,
is defined as when $\MCF(\tau)/\MCF(0)$ crosses the red dashed circle
at a radius of $\e^{-1/2}$. The more wind-blown lag error speckles
remain coherent longer since the AO system repeats the same error
as the wind carries the aberrations across the pupil. This function
and its two main behaviors (straight decorrelation and systematic
phase shift, see Eq.~\ref{eq:MCF-spinningPhase}) determine the loss
of SNR for various science camera exposure times (see Sec.~\ref{sub:Exposure-PCoh}). }
\end{figure}
A more quantitative view of the speckle evolution and decorrelation
is shown in Fig.~\ref{fig:Speckles-MCF-4panel1}. As described by
Eq.~\ref{eq:MCF-spinningPhase}, part of the MCF describes the speckles'
decorrelation, while a phasor factor systematically rotates the phase
depending on the wind's projection onto the selected point's position
relative to the star. The figure shows the MCFs, normalized by the
zero-lag value (note $\MCF(0;\thv)=\PHIsp(k\thv)$), for a single
selected point in the second Airy ring for all of our speckle data
cubes. This complex function shows the speckle decorrelation as a
decrease in the absolute value of $\MCF(\tau)$ with increasing time,
as well as the systematic phase rotation caused by the wind. But also,
the figure shows that when the wind is faster, the MCF falls in magnitude
much more slowly. This is due to the AO system repeating the same
error as it ``chases'' the wind-driven aberrations across the pupil.
While this effect increases speckle noise due to fewer statistically-independent
speckles in a given exposure, it helps us in that longer-lived speckles
can be imaged with more modest science camera frame rates, like 30
fps. Both speckle decorrelation and phase rotation affect our interferometric
measurement's SNR, as we will see below.

\begin{figure}
\centering{}\includegraphics[width=1\columnwidth]{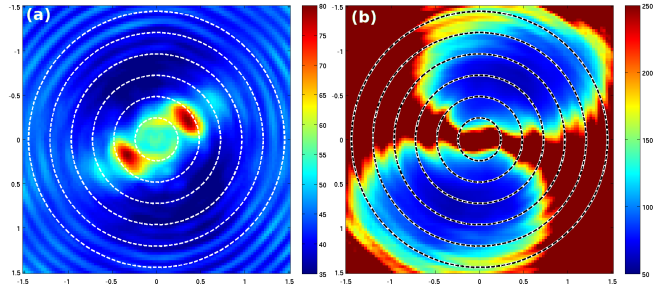}\caption{\label{fig:Halo-tau0}Average speckle halo characteristic timescales
across the focal plane (dashed circles at $1\ld$ spacings). The wind-driven
speckle plume varied about a direction roughly 60 degrees clockwise
from the positive y-axis. Image (a) is the characteristic coherence
time computed using $\left|\Gamma(\tau)/\Gamma(0)\right|=\e^{-1/2}$.
The color scale is labeled in milliseconds and runs from 35~ms to
90~ms. The fitting error speckles typically decorrelated after 35--50~ms
while the lag error speckles in the apparent direction of the wind
lasted longer. Note that the lag-error speckle lifetimes were longer
when they appeared over an Airy ring, which was not expected. Image
(b) shows the average time required for the speckle phase to change
by $\pm\pi/2$, extrapolated from the phase swing after 4 WFS frames.
The color scale runs from 50 to 250~ms. }
\end{figure}
In the pre-AO pupil plane, the coherence time is usually defined in
terms of advection of the wavefront by pure Taylor flow. This implies
that the natural definition of coherence time is $\SF(\wind\tau_{0})=1$,
or as is often quoted, $\tau_{0}=0.314r_{0}/\wind$ for a Kolmogorov
wavefront \citep{greenwood1976power}. However, this is more properly
defined in terms of the temporal MCF, $\Gamma(\tau)=\left\langle \field(\xx+\wind\tau)\field^{*}(\xx)\right\rangle =\exp\left\{ -\SF\left(\wind\tau\right)/2\right\} $,
where the canonical structure function definition becomes $\left|\Gamma(\tnaught)/\Gamma(0)\right|=\e^{-1/2}$.
This is the definition we used to compute the focal plane coherence
time in Fig.~\ref{fig:Halo-tau0}(a). Note that the coherence time
is only about 35--50~ms except in the speckle plume where the coherence
time is much longer. This is the expected behavior with a non-predictive
AO servo algorithm that uses only the most recent WFS measurements
to update the DM. It is interesting to note in Fig.~\ref{fig:Halo-tau0}(a)
that the lag-error speckle coherence time is longer over the Airy
rings. This was not expected, but may be due to the wind bringing
in unseen aberrations at the edge of the pupil. More study is required. 

The other timescale of concern is how long it takes for the systematic
phase shift to change by $\pi/2$, after which at least some of the
interferometric reference beam contributions will start to subtract.
Note that outside of the speckle plume, the phase shift timescale
is longer due to the oblique translation geometry or irrelevant due
to the lack of persistent speckles. For speckles that are more in-line
with the wind flow, the phase change becomes more consistent and the
lag error speckle plume makes the effect dominant. For our data set
however, the most important effect over the majority of the focal
plane is the changing residual aberration pattern, not the translation
by the wind. 

\begin{figure}
\centering{}\includegraphics[width=6cm]{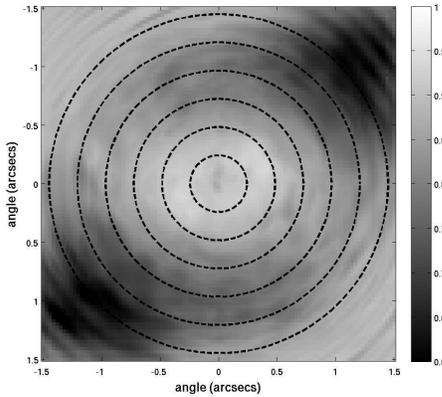}\caption{\label{fig:SNR-effect-MMT}The pixel-by-pixel worst-case loss of SNR
due to science camera exposure time, compared to an ideal short-exposure
SNR. The Clio exposures were 6.5 times longer than the WFS exposures,
so there was some loss of SNR depending on the wind and focal plane
position. The SNR was only down to about 90\% nearer to the star,
dropping to about 60\% in the processing-lag speckle plume as the
phase wrapping rate increased farther from the star. For this bright
star and weather circumstances, we would have no difficulty analyzing
low-order aberrations with even significantly longer exposures.}
\end{figure}

The most useful and practical metric is the additional loss of SNR
due to longer science camera exposures. As expressed in Eq.~\ref{eq:MCF-SNReffect},
the SNR is reduced by a weighted integral over the MCF at various
lags. This can be computed from our WFS data and is shown in Fig.~\ref{fig:SNR-effect-MMT}.
 The figure shows that although the Clio exposure times were long
enough to show a great deal of coherence loss end-to-end, the SNR
dropped to only about 60\% to 90\% of an arbitrarily high frame rate
camera. This can be recovered everywhere, if required, by integrating
about 3 times longer. 

In each of these figures, there are occasional resonant speckles caused
by processing lag and wind, along with the particular characteristics
of the reconstructor, lead to enhanced speckles at certain angles.
This was an unusual situation, possibly due to the engineering tests
being performed. However, such phenomena can occur in a live AO system
and may cause difficulties in the measurement. In this case, the speckle
phase distribution about the mean was still sufficient to give a good
statistical measurement of the static halo. But they are certainly
a sign that greater care should be taken.

\subsection{\label{sub:Static-Halo-Est}Estimating the static halo}

\begin{figure}
\begin{centering}
\includegraphics[width=6cm]{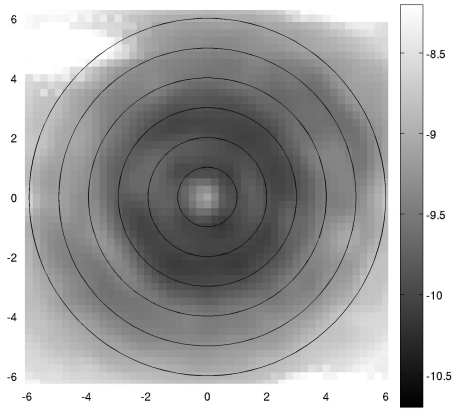}
\par\end{centering}

\centering{}\caption{\label{fig:Gain-map}Map of the scale factor $g(\thv)$ in a neighborhood
around the star. The scale is base-10 log and the circles are 1--6
$\ld$. The white area on the upper left is due to a dithered flaw
in the sensor.}
\end{figure}

\begin{figure}
\centering{}\includegraphics[width=1\columnwidth]{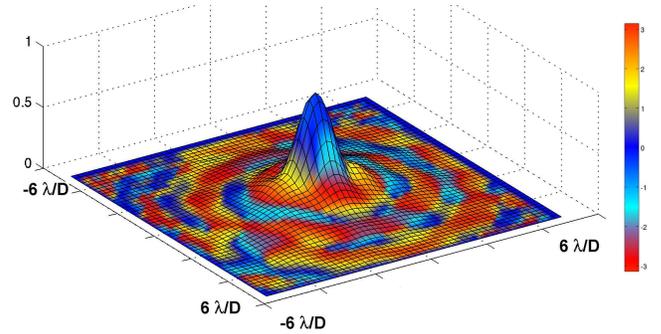}\caption{\label{fig:MMT-StaticHalo-ComplexSurface}Average complex halo derived
from 67 2.5~s Clio image cubes (6700 images) and 43,550 WFS measurements.
This calculation used the presumed ensemble average method for computing
the speckles and therefore may be missing an additive fraction of
the ideal complex halo. }
\end{figure}

As derived in Sec.~\ref{sub:Random-Ref-Interferometry}, we can estimate
the static halo from the Clio and complex speckle data cubes using
Eq.~\ref{eq:TheAnswer}. In discrete form this is 
\begin{equation}
\overline{\Psi}_{nm}=\frac{\sum_{w}\field_{nmw}I_{nmw}}{g_{nm}\sum_{w}\left|\field_{nmw}\right|^{2}}.\label{eq:StaticHalo-est}
\end{equation}
The scale factor is a function of position and is computed using Eq.~\ref{eq:scale-factor},
shown in Fig.~\ref{fig:Gain-map}. As expected for a reconstructor
that is not carefully calibrated, the speckles do not have a uniform
scale over the working field. 
 Including this correction, we now have an estimate of the static
complex halo as shown in Fig.~\ref{fig:MMT-StaticHalo-ComplexSurface}.
Armed with this information, we can now use it in an anti-halo servo
to suppress the halo over some region of interest \citep{CodonaAngel_ApJ_2004},
or we can use the entire halo to estimate the wavefront in the pupil
plane including the NCP error to improve the Strehl ratio and imaging
quality at the science camera. This use of the measured static halo
requires more work but is conceptually as simple as taking the Fourier
transform of the static halo.

\subsection{\label{sub:Pupil-Field-Est-with-NCP}Estimating the pupil field}

\global\long\def\dz{\delta z}
The static halo is the mean of the complex halo (App.~\ref{app:additive_speckles},
specifically Eq.~\ref{eq:Halo}), and therefore it is the inverse
Fourier transform of the mean complex pupil field $\pfield(\xx,t)$
under the paradigm of Fourier optics (Eq.~\ref{eq:PFieldBar_from_staticHalo}).
Our static halo field includes the effects of any non-common-path
aberrations after the AO system and the science camera, which may
have been introduced at any point in the intervening light path. The
inverse Fourier transform of the static halo does not give the literal
pupil field, but the apparent pupil field including the effects of
the downstream aberrations back-projected to the pupil. Since the
pupil is the canonical location of the DM where we can affect wavefront
corrections, we can use this estimate to determine the wavefront offsets
required to modify the resulting halo in any desired way. Expressing
this in terms of the forward Fourier optics equation, Eq.~\ref{eq:Fourier-Optics},
we write 
\begin{equation}
\pfield_{0}(\xx)\pupil(\xx)=\frac{\int\e^{-i\kkv\cdot\xx}\overline{\halo}(\kkv)\dd^{2}\kappa}{\left(2\pi\right)^{2}\SR^{1/2}}\label{eq:Pupil-field-with-NCP1}
\end{equation}
where we have made use of the assumption that the residual wavefront
phase from the AO correction is a zero-mean Gaussian random variable
and $\left\langle \exp(i\phase)\right\rangle =\exp(-\left\langle \phase^{2}\right\rangle /2)\equiv\SR^{1/2}$
is the square root of the Strehl ratio. This gives us an estimate
of the pupil field, blurred due to the limited spatial frequencies
recovered by the interferometry calculation. The NCP pupil field is
$\bar{\field}(\xx)=a(\xx)\exp\left(ik\dz\right)$ where the amplitude
$a(\xx)\in\REAL$ may include transmission effects, as well as back-projections
of downstream vignetting, etc. The wavefront to be corrected is given
by 
\begin{equation}
\dz=\frac{\lambda}{2\pi}\arg\left\{ \int\e^{-i\kkv\cdot\xx}\overline{\halo}(\kkv)\dd^{2}\kappa\right\} .\label{eq:NCP-WFE-estimate}
\end{equation}

This simple procedure should be all that is required to compute the
wavefront correction, but there are some caveats and corrections.
The pupil field estimate is missing some low spatial frequency components,
mostly a constant. The reasons for this have already been discussed
in Sec.~\ref{sec:Simulation}, and are related to our computed phase
referencing to the PSF core, and the fact out tip-tilt stabilization
reducing the apparent effect of anti-Hermitian speckle pairs near
the optical axis. We also did not account for interference between
the very low spatial frequency speckle pairs, causing an error that
becomes significant within $\lambda/2D$ of the PSF core. We can calibrate
this error by applying a known bias to the DM and seeing what the
measured value is. Since a constant bias would affect the measured
phase aberration according to 
\begin{equation}
\widehat{\dz}=\frac{\lambda}{2\pi}\arg\left\{ \field_{bias}+\int\e^{-i\kkv\cdot\xx}\overline{\halo}(\kkv)\dd^{2}\kappa\right\} \label{eq:LFBiased-WFE-estimate}
\end{equation}
we can then adjust the constant $\field_{bias}$ to give a reasonable
estimate of the actual aberration.

\begin{figure}
\centering{}\includegraphics[width=1\columnwidth]{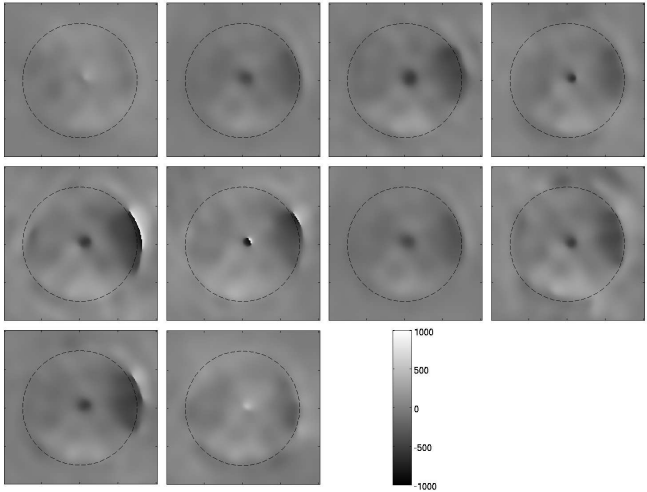}\caption{\label{fig:Pupil-Wavefront-shortExamples}Examples of 2.5~s estimates
of the pupil wavefront. The scale factor $g(\thv)$ was computed from
the full set of 67 100-exposure image cubes and used in the calculation
of the static halo for each cube. The loss of low spatial frequency
power due to the inability to sense piston as well as tip-tilt stabilization
was approximately compensated for by adding a constant to the computed
pupil field. The resulting phase was converted to mirror surface heights
using $z=\phi(\xx)\lambda/4\pi.$ The scale is labeled in nanometers.}
\end{figure}

\begin{figure}
\centering{}\includegraphics[width=1\columnwidth]{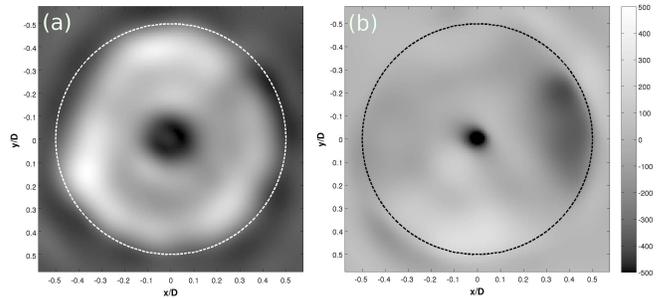}\caption{\label{fig:Biased-Pupil-Wavefront-Avg}The MMT's average static pupil
field including back-projected non-common-path aberrations. The pupil
field was biased by a constant of 1/2 the rms amplitude to approximately
compensate for underestimated low-spatial-frequency power. Image (a)
is the intensity, and the $0.1D$ central obstruction is clearly seen.
Image (b) is the average WFE computed as a bias to be applied to the
DM. The scale is labeled in nanometers. The peak-to-valley surface
error is 1266~nm and the rms error is 206~nm (an rms phase error
of $\lambda/9.2$ at L band). }
\end{figure}

Since we did not have a calibration for the data presented in this
paper, we estimated a constant pupil field of 50\% of the mean computed
value. Adding this to the computed value for each 2.5~s Clio data
cube, we compute the NCP aberrations. A selected subset of the wavefront
errors are shown in Fig.~\ref{fig:Pupil-Wavefront-shortExamples}.
 Since the estimates are generated from only 2.5 s duration science
camera cubes, the recovered NCPs show a significant amount of variance,
but they also show consistent morphological patterns. The average
of these is shown in Fig.~\ref{fig:Biased-Pupil-Wavefront-Avg}.
 The peak-to-valley surface error (i.e.~half the wavefront error)
is 1266~nm and the rms error is 206~nm. This result is compatible
with the science images, and all that remains is to apply the corrections
to the DM and see the improvement in the image quality.

\section{Discussion\label{sec:discuss}}

We have developed a new interferometric technique for focal plane
wavefront sensing using the residual starlight speckles left behind
by an AO system. Our method requires no extra hardware or optics beyond
that used in a typical AO system. The rapidly-changing residual AO
speckles are present in the science camera image plane and coherently
interfere with the starlight diffraction halo. The speckles result
from uncontrolled errors in the correction of the atmospheric aberrations,
resulting in small residual wavefront fluctuations in a pupil plane
downstream from the AO system's deformable mirror (DM). While uncontrolled
and unsuppressed by the AO system, these residual errors are still
monitored by the wavefront sensor and are available for other purposes.
Improved AO technology and algorithms will continually reduce the
errors and subsequent speckles, but they will remain a ubiquitous
feature that can be exploited for other purposes. Our technique uses
the wavefront sensor measurements to compute a numerical analogue
of the complex halo (with both amplitude and phase), which is also
measured in intensity by the science camera. The computed halo is
used as a key to interpret short-exposure science camera images, allowing
the steady portion of the star's complex halo. Since the intensity
measurements are downstream from the AO system, the our complex halo
measurement includes information about both common and non-common-path
aberrations. This information is useful both for correcting the wavefront
before imaging and post-detection image processing. 

We demonstrated the technique using data taken with the MMT AO wavefront
sensor (WFS) and the Clio science camera in the L band. We synchronized
and down-sampled the WFS measurements to match the Clio frame rate
of about 40 frames per second. Using our mathematical theory, we were
able to compute an estimate of the complex halo every 2.5~s from
100 Clio images and 650 WFS measurements. The resulting complex halo
estimate represents the average field in the focal plane and is related
to the pupil field by Fourier transform. We analyzed the temporal
statistics of the speckle halo and used them to estimate the effect
of slower science camera exposures. We found that the impact of exposure
time, while depending on the details of wind and focal plane position,
was not serious---especially near the star where halo suppression
will be most important.

Although we were not able at this time to demonstrate the use of
the complex halo to improve the science images, the path forward and
potential benefits of continuous end-to-end optical quality measurements
and updating are clear. The Fourier transform of the halo provides
an estimate of the pupil field, including the back-propagated effects
of NCP aberrations. By applying appropriate DM offsets to compensate
for the static pupil field's phase, the Strehl ratio seen by the science
camera can be improved, providing correspondingly improved detection
sensitivity. Measurements of the complex halo are even more directly
useful for suppressing residual diffracted starlight in a flawed or
poorly-aligned coronagraph by using an antihalo servo, the theory
of which was not discussed in this paper. 

In future work we plan on developing a system to estimate the static
halo in real time during observations. The estimates will then be
fed back to the AO system either as WFS or DM biases. For normal (non-coronagraphic)
imaging, the goal will be to optimize Strehl ratio at the science
camera. When using a coronagraph, our goal will be to suppress unwanted
halo in the search region while simultaneously maximizing Strehl ratio.
These enhancements will not require the addition of any new optics,
but will require the addition of an outer control loop that affects
the operation of the AO system. We also plan to use the complex halo
measurements to develop two new image processing algorithms. By using
the estimated pupil field, we can compute better numerical PSFs, which
include the effects of both common and non-common-path aberrations
as well as optical filter bandwidth. Subtracting the computed PSFs
from the actual star images will reduce speckle noise. We also plan
to use the computed PSFs as a key for when to keep pixels in the science
data cube, including them only when the computed halo falls below
some statistical threshold. This ``lucky pixel'' type algorithm
should be capable of increasing sensitivity by a factor of about 2.
Including both algorithms, the decrease in speckle noise can be significant,
depending on the raw performance of the AO system. For the MMT, we
would be hoping to achieve an additional sensitivity of 2--4 magnitudes
over the corrected coronagraph alone.

\appendix{}

\section{The static halo and speckle cloud}

\label{app:additive_speckles}

We treat residual AO speckles to be additive random field fluctuations
coherently interfering with a static diffraction pattern. In this
appendix we explore the halo, and consider assumptions and statistics.
We show why the high Strehl ratio focal plane field is a somewhat
attenuated version of the telescope's diffraction pattern, including
non-common-path aberrations, and a coherent cloud of speckles which
are reasonably described as zero-mean Gaussian complex field fluctuations.
We will consider two cases: the classic seeing-limited situation with
no AO wavefront correction as a touchstone, and the diffraction-limited
case with a static PSF surrounded by a cloud of speckles. 

We start with an incident field $\pfield(\xx)$ passing through a
possibly complex pupil mask $\pupil(\xx)$, which also includes any
non-common-path (NCP) aberrations. We will not include scintillation
here, but could easily do so. For convenience and wavelength independence,
we use the spatial frequency $\kk=(\kappa_{x},\kappa_{y})$ to describe
the complex halo field
\begin{equation}
\halo(\kk,t)=\int\e^{i\kk\cdot\xx}\pfield(\xx)\pupil(\xx)\dd^{2}x\label{eq:Halo}
\end{equation}
where 
\[
\pfield(\xx)=\e^{i\pphase(\xx,t)}\pfield_{0}(\xx)
\]
is the pupil field and $\pfield_{0}(\xx)$ is the pupil field without
any of the fluctuating phase aberrations $\pphase(\xx,t)$. Note that
our halo field is the \emph{angular spectrum} of plane waves comprising
the pupil field, where each $\kk$ plane wave travels from the direction
$\thv=\kk/k$ with $k=2\pi/\lambda$. The \emph{point spread function}
(PSF) is the focal plane intensity distribution resulting from an
on-axis star, where $\pfield_{0}(\xx)=1$. Although non-common-path
aberrations occur downstream from the AO system and the nominal pupil
plane, our small field of interest allows us to treat them as being
included in the initial pupil field $\pfield_{0}(\xx)$. In our notation,
the PSF is the \emph{angular power spectrum} of plane waves

\begin{equation}
\PSF(\kk,t)=\left|\halo(\kk,t)\right|^{2}.\label{eq:PSF}
\end{equation}

The fluctuating phase aberrations are presumed to vary randomly in
space and time, with ergodicity allowing us to use ensemble and temporal
averages interchangeably. While this may be a reasonable assumption
for the atmospherically-distorted incident field, it is wise to keep
an open mind regarding the post-AO statistics. In general, the average
PSF is 

\begin{equation}
\left\langle \PSF(\kappa)\right\rangle =\int\dd^{2}x_{1}\int\dd^{2}x_{2}\e^{i\kk\cdot(\xx_{1}-\xx_{2})}\left\langle \e^{i(\pphase(\xx_{1},t)-\pphase(\xx_{2},t))}\right\rangle \pfield_{0}(\xx_{1})\pfield_{0}^{*}(\xx_{2})\pupil(\xx_{1})\pupil^{*}(\xx_{2}).\label{eq:genAvgPSF}
\end{equation}
The average phase exponential is traditionally simplified using 
\begin{equation}
\left\langle \e^{iq}\right\rangle =\e^{-\left\langle q^{2}\right\rangle /2}
\end{equation}
which is valid when $q$ is a zero-mean Gaussian random variable (GRV).
If we assume $q=\pphase(\xx_{1})-\pphase(\xx_{2})$ is a GRV, then
\begin{equation}
\left\langle \e^{i(\pphase(\xx_{1})-\pphase(\xx_{2}))}\right\rangle =\e^{-D_{\pphase}(\xx_{1},\xx_{2})/2}\label{eq:birth-of-the-SF}
\end{equation}
where 

\begin{equation}
D_{\pphase}(\xx_{1},\xx_{2})=\left\langle (\pphase(\xx_{1})-\pphase(\xx_{2}))^{2}\right\rangle 
\end{equation}
is the structure function of the pupil field phase. Again, while Eq.~\ref{eq:birth-of-the-SF}
can reasonably be applied to the seeing-limited case, the post-AO
case may contain patterns and correlations that break this assumption.
This may be particularly important when using this result with multi-segment
telescopes where the segments are being phased with a slower servo
loop, etc. We cannot address that case here, just provide a warning.
The statistics of the uncorrected atmospheric phase are also usually
considered to be translation independent, while the post-AO statistics
may well not be; post-AO correction residuals vary depending on position
relative to actuators, WFS sub-apertures, pupil edges, etc. A simple
example is the mean-square phase residual, $\left\langle \pphase^{2}(\xx)\right\rangle =\sigma_{\pphase}^{2}$
, which is generally assumed to be position independent, although
it probably is not. We note but ignore this for now, assuming that
at least the second-order phase statistics are translation independent.
If so, then 
\[
D_{\pphase}(\xx_{1},\xx_{2})\rightarrow D_{\pphase}(\xx_{1}-\xx_{2}).
\]
Because the post-AO halo contains lag error speckles, we cannot generally
assume isotropy since the wind defines a preferred direction. Translation
independence allows us to simplify Eq.~\ref{eq:genAvgPSF}. Make
the coordinate change 
\begin{eqnarray*}
\alphav & = & \left(\xx_{1}+\xx_{2}\right)/2\\
\betav & = & \xx_{2}-\xx_{1}
\end{eqnarray*}
and carry out the $\alphav$ integration. The result gives the average
PSF in terms of a spatially-filtered version of the \emph{optical
transfer function} (OTF) 

\begin{equation}
\left\langle \PSF(\kappa)\right\rangle =\int\e^{i\kk\cdot\betav}\e^{-D_{\pphase}(\betav)/2}\OTF(\betav)\dd^{2}\beta\label{eq:meanPSF}
\end{equation}
where the OTF \citep{IntroOTF_Williams_Becklund_SPIE2002} is given
by

\begin{equation}
\OTF(\betav)=\int\pfield_{0}(\alphav+\betav/2)\pupil(\alphav+\betav/2)\pfield_{0}^{*}(\alphav+\betav/2)\pupil^{*}(\alphav+\betav/2)\dd^{2}\alpha.\label{eq:basicOTF}
\end{equation}
Since $D_{\pphase}(\betav)$ starts off at zero when $\betav=0$ and
rises to a maximum value as $\left\Vert \betav\right\Vert $ increases,
Eq.~\ref{eq:meanPSF} shows how the mean PSF loses higher angular
frequencies as the rms pupil wavefront phase increases. This expression
will be our guide in dissecting the halo, both for the intensity PSF
and realizations of the complex halo itself.

The uncorrected structure function has certain common features that
we often refer to. As two phase measurement points become more distantly
separated, the phase values eventually become completely uncorrelated
and 

\begin{equation}
D_{\pphase}(\ss)\rightarrow2\sigma_{\pphase}^{2}.\label{eq:SF-asymptote}
\end{equation}
In the post-AO case, this asymptotic limit applies as $s=\left\Vert \ss\right\Vert \gg\lao$
where $\lao$ is the correlation scale for the AO system, defined
below. In the no-AO seeing-limited case, this does not apply until
the spacing exceeds at least the outer scale $\louter$, and possibly
not until significantly beyond that, depending on definitions. It
is not uncommon to consider a case, albeit non-physical, where the
outer scale and the phase variance are infinite. For simplicity in
this discussion we will assume that there is an outer scale and $\left\langle \pphase(\xx)\right\rangle =0$.

The no-AO case is often modeled, or at least described, using the
isotropic Kolmogorov power law structure function, parametrized by
the Fried length $r_{0}$, 
\begin{equation}
D_{\phase}(s)=6.88\left(\frac{s}{r_{0}}\right)^{5/3}.\label{eq:Kolmogorov}
\end{equation}
This function actually flattens out past $\louter$, but for $D/r_{0}\gg1$
it will not change this discussion. The blurring factor in Eq.~\ref{eq:meanPSF},
$\exp\left\{ -D_{\pphase}(\beta)/2\right\} $, has a radius defined
by $D_{\pphase}(\beta)/2=1$ or $\beta=0.477r_{0}$. For our estimates,
we will use the diameter of the blurring filter as $\sim r_{0}$.
The radius of the OTF is the pupil diameter $D$ , although it may
be more complicated for a large segmented telescope pupil. If $D/r_{0}\gg1$,
the blurring filter limits the integral in Eq.~\ref{eq:meanPSF}
and we can approximate the OTF with its value at the origin, which
is simply the area of the pupil weighted by its transmission mask.
Therefore in the seeing-limited no-AO case, the average PSF is approximately
given by 
\begin{equation}
\left\langle \PSF(\kappa)\right\rangle \approx\OTF(0)\int\e^{i\kk\cdot\betav}\e^{-D_{\pphase}(\betav)/2}\dd^{2}\beta.\label{eq:noAO-PSF}
\end{equation}

When the telescope is equipped with a good AO system, the residual
phase structure function has a maximum that is still relatively small,
causing the blurring factor to look like the constant $\exp\left\{ -\sigma_{\pphase}^{2}\right\} $
with a small rise up to the value of 1 as $\betav\rightarrow0$. This
no longer limits the integral in Eq.~\ref{eq:meanPSF} and reveals
a diffraction-limited PSF plus a scattered light halo. This effect
of this can best be seen by explicitly pulling out the constant term
and letting the small rise near the origin stand alone 
\begin{eqnarray}
\e^{-D_{\pphase}(\betav)/2} & = & \SR+(1-\SR)H(\betav)\label{eq:partition-expSF}
\end{eqnarray}
where 
\begin{equation}
\SR=\e^{-\left\langle \pphase^{2}\right\rangle }
\end{equation}
is the Strehl ratio as usually expressed in Maréchal's approximation
\citep{Hardy98}, and the factor 
\begin{equation}
H(\betav)=\frac{\e^{-D_{\pphase}(\betav)/2}-\SR}{1-\SR}\label{eq:AO-blurring}
\end{equation}
goes from a maximum value of $H=1$ at $\betav=0$ to $H\rightarrow0$
for $\left\Vert \betav\right\Vert >\lao$. Just as the structure function
exponential limited the integral (Eq.~\ref{eq:meanPSF}) in the seeing-limited
case, $H(\betav)$ limits the integral in the scattered halo term
when $D/\lao\gg1$. This gives a high-Strehl approximation for the
PSF as
\begin{equation}
\left\langle \PSF(\kappa)\right\rangle \approx\SR\PSF_{0}(\kappa)+\left(1-\SR\right)\OTF(0)\widetilde{H}(\kk)\label{eq:AO-PSF}
\end{equation}
where
\begin{equation}
\widetilde{H}(\kk)=\int\e^{i\kk\cdot\betav}H(\betav)\dd^{2}\beta
\end{equation}
and 
\begin{equation}
\PSF_{0}(\kappa)=\int\e^{i\kk\cdot\betav}\OTF(\betav)\dd^{2}\beta
\end{equation}
is the diffraction-limited PSF (including NCP effects since it includes
$\pfield_{0}(\xx)$). Note that Maréchal's approximation is always
true, but only for the diffraction-limited part of the PSF. 

So in both the low-Strehl seeing-limited case Eq.~\ref{eq:noAO-PSF}
and the high-Strehl post-AO case Eq.~\ref{eq:AO-PSF}, we have a
spatial filter that limits the $\beta$ integral in Eq.~\ref{eq:meanPSF}.
In one case the spatial filter diameter is $r_{0}$, while in the
other it is $\lao$, the width of $H(\betav)$. Although they were
derived somewhat differently, Eq.~\ref{eq:AO-PSF} actually applies
to both the AO and no-AO cases, corresponding to the high-Strehl and
low-Strehl cases respectively.

Eq.~\ref{eq:AO-PSF} describes the mean high Strehl ratio PSF as
the sum of a static PSF plus an averaged speckle halo. This is useful
as a description of the mean starlight intensity, and the derivations
are clear and familiar since the product of the field and its conjugate
lead to phase differences and hence phase structure functions that
are insensitive to large-scale phase wander. The same is not true
of the field, which has an overall phase that possibly wanders over
many $2\pi$ cycles. Even a highly-corrected post-AO field can wander
in phase since the wavefront sensor and any corrected images are blind
to the phase. If we somehow measured the complex halo field directly
while this large uncontrolled piston phase wander was occurring, the
mean halo field would tend to zero. We get around this problem by
using a phase reference characteristic of the field over the pupil
as a whole, $\phaseref(t)$. The phase-referenced halo field is then
\[
\halo(\kk,t)\rightarrow\halo(\kk,t)\e^{-i\phaseref(t)}.
\]
In our interferometric analysis, we compute the wavefront from the
WFS measurements and then use an optical model and Fourier optics
to compute an estimate of the halo field. We tip-tilt stabilize the
PSF images and remove the mean slopes from the WFS measurements, so
the phase of the PSF core is equivalent the halo field on the optical
axis. (Note that if we were modeling a coronagraph, the corresponding
phase reference would be defined by the PSF core before encountering
the focal plane mask.) The reference phase is therefore 
\begin{equation}
\phaseref(t)=\arg\left\{ \int\e^{i\pphase(\xx,t)}\pupil(\xx)\dd^{2}x\right\} \label{eq:RefPhase_PSFCore}
\end{equation}
which fluctuates somewhat depending on the residual phase pattern.
We can now write the mean phase-referenced pupil field
\begin{equation}
\overline{\pfield}(\xx)=\left\langle \e^{i\left(\pphase(\xx,t)-\phaseref(t)\right)}\right\rangle \pfield_{0}(\xx).\label{eq:meanPfield0}
\end{equation}
We have already made use of the average of a phasor with a GRV phase
when we found the phase structure function in Eq.~\ref{eq:birth-of-the-SF}.
Since $D/\lao\gg1$, many statistically-independent regions of the
pupil contribute to the integral in Eq.~\ref{eq:RefPhase_PSFCore},
giving a result that is at least approximately a GRV. Assuming further
that the AO correction is at least fairly good, we can assume that
$\sigma_{\pphase}<\pi/2$ or better. In very high Strehl, the integral
is many sigmas from the origin and the complex argument approaches
a projection, giving a Gaussian distribution for the reference phase.
Its scatter will also presumably be much less than $\sigma_{\pphase}$,
by a factor of $\sim\lao/D$ based on the number of independent contributing
patches. Therefore, we can argue that $\phaseref(t)$ is a GRV and
uncorrelated with any given $\pphase(\xx,t)$ by at least the area
factor of $(\lao/D)^{2}$. With these assumptions, the mean pupil
field becomes
\begin{equation}
\overline{\pfield}(\xx)=\e^{-\left\langle \pphase^{2}\right\rangle /2}\e^{-\left\langle \pphase_{ref}^{2}\right\rangle /2}\pfield_{0}(\xx)=\e^{-\left\langle \pphase_{ref}^{2}\right\rangle /2}\SR^{1/2}\pfield_{0}(\xx).\label{eq:meanPfield0-1}
\end{equation}
If the Strehl ratio is high, as it is when working with corrected
MMT images in the mid-infrared, we can neglect the reference phase
variance, giving us the mean halo formula we use in our analysis
\begin{equation}
\overline{\pfield}(\xx)=\SR^{1/2}\pfield_{0}(\xx).\label{eq:simpleMeanPfield}
\end{equation}
Carrying out the same phase referencing and averaging on Eq.~\ref{eq:Halo}
with high-Strehl average pupil field, Eq.~\ref{eq:simpleMeanPfield},
we can inverse Fourier transform to find the mean pupil field from
the mean (or \emph{static}) phase-referenced halo field 
\begin{equation}
\pfield_{0}(\xx)\pupil(\xx)=\frac{\int\e^{-i\kk\cdot\xx}\overline{\halo}(\kk)\dd^{2}\kappa}{(2\pi)^{2}\SR^{1/2}}.\label{eq:PFieldBar_from_staticHalo}
\end{equation}
Therefore, once we have an estimate of the static halo, we can use
Eq.~\ref{eq:PFieldBar_from_staticHalo} to estimate the actual pupil
field, which includes the NCP aberrations. Since the Strehl ratio
is real, the NCP phase aberrations are independent of the Strehl ratio
and are given by
\begin{equation}
\pphase_{NCP}(\xx)=\arg\left\{ \int\e^{-i\kk\cdot\xx}\overline{\halo}(\kk)\dd^{2}\kappa\right\} .\label{eq:NCP_phase_from_staticHalo}
\end{equation}

The pupil field is the mean plus the zero-mean residuals, which contribute
to the focal plane speckles as described above. The speckle field
at any given instant is the sum of the complex contributions from
each statistically independent patch, rotated in the complex plane
by the local Fourier kernel. If $D/\lao\gg1$, there are many independent
patches, leading to zero-mean complex Gaussian speckle fields, even
though the pupil field is far from Gaussian.

\section{Gaussian Complex Random Variables}

\label{app:GRV}We use some less familiar properties of complex Gaussian
random variables in this paper and to assist the reader, we will provide
some derivations. We consider an ensemble of complex values $\psi_{n}\in\COMPLEX$
with a Gaussian, or ``normal'' distribution. The Gaussian distribution
has only two independent moments: the mean $\mu=\left\langle \psi\right\rangle $
and the standard deviation $\sigma=\left\langle \left|\psi-\left\langle \psi\right\rangle \right|^{2}\right\rangle ^{1/2}$.
For simplicity, we will limit the discussion here to isotropic distributions
(i.e. where the standard deviation depends on direction in the complex
plane), but the properties we require generalize to non-isotropic
distributions as well. The isotropic probability distribution is 
\begin{equation}
\Pr\left(\psi\right)=\frac{\e^{-\left|\psi-\mu\right|^{2}/2\sigma^{2}}}{2\pi\sigma^{2}}.\label{eq:CGaussianProbDist}
\end{equation}
As usual, if $\psi=\chi+i\zeta=\rho\e^{i\vartheta}+\mu$ , $(\chi,\zeta,\rho,\vartheta)\in\REAL$,
the probability distribution is normalized 
\begin{equation}
\int_{-\infty}^{\infty}\int_{-\infty}^{\infty}\Pr(\psi)\dd\chi\dd\zeta=1\label{eq:GRV_norm}
\end{equation}
and the ensemble average of a quantity is computed by integrating
the quantity weighted by the probability distribution 
\begin{equation}
\left\langle f(\psi)\right\rangle \equiv\int_{-\infty}^{\infty}\int_{-\infty}^{\infty}f(\psi)\Pr(\psi)\dd\chi\dd\zeta.\label{eq:GRV_mean_func}
\end{equation}
This gives the mean as
\begin{equation}
\mu=\int_{-\infty}^{\infty}\int_{-\infty}^{\infty}\psi\Pr(\psi)\dd\chi\dd\zeta\label{eq:GRV_mean}
\end{equation}
and the variance
\begin{equation}
\sigma^{2}\equiv\int_{-\infty}^{\infty}\int_{-\infty}^{\infty}\left|\psi-\mu\right|^{2}\Pr(\psi)\dd\chi\dd\zeta.\label{eq:GRV_var}
\end{equation}

The lesser known results that we require are $Y=\left\langle \psi^{2}\right\rangle $
(as opposed to the more important $X=\left\langle \psi\psi^{*}\right\rangle $)
and the third moment $Q=\left\langle \psi^{2}\psi^{*}\right\rangle $.
For our purposes, $\psi$ is zero-mean, which somewhat simplifies
the derivations here. We wish to prove that the ensemble average of
both $Y$ and $Q$ are zero, and that an $N$-sample estimate of each
has a scatter that is proportional to $1/\sqrt{N}$. 

Using eq.~\ref{eq:GRV_mean_func} we can write
\[
\]
\begin{eqnarray}
Y=\left\langle \psi^{2}\right\rangle  & = & \int_{-\infty}^{\infty}\int_{-\infty}^{\infty}\psi^{2}\Pr(\psi)\dd\chi\dd\zeta\label{eq:GRV_Ymoment_deriv}\\
 & = & \int_{-\infty}^{\infty}\int_{-\pi}^{\pi}\rho^{2}\e^{2i\vartheta}\frac{\e^{-\rho^{2}/2\sigma^{2}}}{2\pi\sigma^{2}}\rho\dd\vartheta\dd\rho\\
 & = & \underbrace{\left(\int_{-\pi}^{\pi}\e^{2i\vartheta}\dd\vartheta\right)}_{=0}\int_{-\infty}^{\infty}\rho^{3}\frac{\e^{-\rho^{2}/2\sigma^{2}}}{2\pi\sigma^{2}}\dd\rho.
\end{eqnarray}
Therefore 
\begin{equation}
Y\equiv\left\langle \psi^{2}\right\rangle =0.\label{eq:GRV_Yavg}
\end{equation}
Similarly, 
\begin{eqnarray}
Q=\left\langle \psi^{2}\psi^{*}\right\rangle  & = & \int_{-\infty}^{\infty}\int_{-\infty}^{\infty}\psi^{2}\psi^{*}\Pr(\psi)\dd\chi\dd\zeta\label{eq:GRV_mean_third_deriv}\\
 & = & \int_{-\infty}^{\infty}\int_{-\pi}^{\pi}\rho^{3}\frac{\e^{-\rho^{2}/2\sigma^{2}}}{2\pi\sigma^{2}}\e^{-i\vartheta}\rho\dd\vartheta\dd\rho\\
 & = & \underbrace{\left(\int_{-\pi}^{\pi}\e^{-i\vartheta}\dd\vartheta\right)}_{=0}\int_{-\infty}^{\infty}\rho^{4}\frac{\e^{-\rho^{2}/2\sigma^{2}}}{2\pi\sigma^{2}}\dd\rho\\
Q & = & 0.\label{eq:Q3_ensembleAvg}
\end{eqnarray}
Therefore, for zero-mean isotropic Gaussian distributions, both moments
are indeed zero. 

We also wish to know how the terms converge to zero in the discrete
case with $N$ measurements. An experiment producing $N$ samples
of $\psi$ allows us to make an estimate of the $Y$ moment by averaging
the squares of the individual measurements 
\begin{equation}
Y_{N}=\left\langle \psi^{2}\right\rangle _{N}\equiv\frac{1}{N}\sum_{n=1}^{N}\psi_{n}^{2}.
\end{equation}
We can calculate the scatter in this measurement by considering an
ensemble of the above $N$-measurement experiments, and computing
the variance of $Y_{N}$. The ensemble average of $Y_{N}$ is obviously
zero from Eq.~\ref{eq:GRV_Yavg}. However, the variance of $Y_{N}$
is not zero and is given by
\begin{eqnarray}
\sigma_{Y}^{2} & = & \left\langle \left|Y_{N}-\left\langle Y_{N}\right\rangle \right|^{2}\right\rangle =\left\langle \left|Y_{N}\right|^{2}\right\rangle \\
 & = & \frac{1}{N^{2}}\left\langle \left|\sum_{n=1}^{N}\psi_{n}^{2}\right|^{2}\right\rangle =\frac{1}{N^{2}}\left\langle \sum_{n=1}^{N}\sum_{m=1}^{N}\psi_{n}^{2}\psi_{m}^{*2}\right\rangle .
\end{eqnarray}
If all of the samples $\psi_{n}$ and $\psi_{m}$ in a given experiment
are statistically independent and zero mean, then for $n\ne m$, the
ensemble average of the ``off-diagonal'' terms are zero since $\left\langle \psi_{n}^{2}\psi_{m}^{*2}\right\rangle =\left\langle \psi_{n}^{2}\right\rangle \left\langle \psi_{m}^{2}\right\rangle ^{*}$
and each of the ensemble average $Y$ moment factors are $0$ from
Eq.~\ref{eq:GRV_Yavg}). The ``diagonal'' terms where $n=m$ are
the same sample and do not average to zero. We continue 
\begin{eqnarray}
\sigma_{Y}^{2} & = & \frac{1}{N^{2}}\left\langle \sum_{n=1}^{N}\left|\psi_{n}\right|^{4}\right\rangle \\
 & = & \frac{1}{N^{2}}\sum_{n=1}^{N}\left\langle \left|\psi_{n}\right|^{4}\right\rangle \\
 & = & \frac{1}{N}\left\langle \left|\psi\right|^{4}\right\rangle 
\end{eqnarray}
where we have assumed that all of the measurements have the same statistical
moments. Therefore 
\begin{equation}
\sigma_{Y}=\left\langle \left|\psi\right|^{4}\right\rangle ^{1/2}/\sqrt{N}.\label{eq:sigma_Y}
\end{equation}

Similarly, an $N$-term estimate of $Q$ is
\begin{equation}
Q_{N}=\left\langle \psi^{2}\psi^{*}\right\rangle _{N}\equiv\frac{1}{N}\sum_{n=1}^{N}\psi_{n}^{2}\psi_{n}^{*}.
\end{equation}
The variance of $Q_{N}$ is 
\begin{eqnarray}
\sigma_{Q}^{2} & = & \left\langle \left|Q_{N}-\left\langle Q_{N}\right\rangle \right|^{2}\right\rangle =\left\langle \left|Q_{N}\right|^{2}\right\rangle \\
 & = & \frac{1}{N^{2}}\left\langle \left|\sum_{n=1}^{N}\psi_{n}^{2}\psi_{n}^{*}\right|^{2}\right\rangle \\
 & = & \frac{1}{N^{2}}\left\langle \sum_{n=1}^{N}\sum_{m=1}^{N}\psi_{n}^{2}\psi_{n}^{*}\psi_{m}^{*2}\psi_{m}\right\rangle .
\end{eqnarray}
Once again, so long as the measurements for $n\ne m$ are statistically
independent, so the off-diagonal terms vanish due to Eq.~\ref{eq:Q3_ensembleAvg}.
The diagonal terms remain giving us 
\begin{eqnarray*}
\sigma_{Q}^{2} & = & \frac{1}{N^{2}}\left\langle \sum_{n=1}^{N}\left|\psi_{n}\right|^{6}\right\rangle \\
 & = & \frac{1}{N^{2}}\sum_{n=1}^{N}\left\langle \left|\psi_{n}\right|^{6}\right\rangle \\
 & = & \frac{N}{N^{2}}\left\langle \left|\psi\right|^{6}\right\rangle =\frac{1}{N}\left\langle \left|\psi\right|^{6}\right\rangle .
\end{eqnarray*}
Therefore 
\begin{equation}
\sigma_{Q}=\left\langle \left|\psi\right|^{6}\right\rangle ^{1/2}/\sqrt{N}.\label{eq:sigma_Q}
\end{equation}
The same convergence result holds for any ``unbalanced'' moment
where the number of straight and conjugate factors are unequal.

\acknowledgements{Observations reported here were made at the MMT, a joint facility
of the University of Arizona and the Smithsonian Institution. This
effort was supported by the National Science Foundation grants AST-0804586
and AST-0904839. We would also like to make a special thanks to Roger
Angel for stepping in at a critical point in preparing the NSF proposal
when the PI (JLC) was in the hospital. This project would have, at
the minimum, been significantly delayed had he not taken extraordinary
action. We thank Vidhya Vaitheeswaran for PCR code which helped organise
the synchronization mode required for this data. We thank Phil Hinz
for support with the Clio mid-IR camera. We also thank Michael Hart
and Rich Frazin for review and editorial comments.}

\bibliographystyle{apj}

\end{document}